\documentclass[11pt,a4paper,nofootinbib]{revtex4}

\usepackage{dcolumn}   
\usepackage{bm}        
\usepackage{multirow}
\usepackage{xcolor}

\usepackage{amsmath}
\usepackage{amsfonts}
\usepackage{amssymb}
\usepackage{makeidx}
\usepackage{graphicx}
\usepackage{anysize}
\usepackage{hyperref}
\usepackage{slashed}

\usepackage[title]{appendix}

\usepackage{mathrsfs}

\usepackage{ulem} 

\usepackage{afterpage}

\newenvironment{mymathbox}
{\par\smallskip\centering\begin{lrbox}{0}%
\begin{minipage}[c]{0.8\textwidth}}
{\end{minipage}\end{lrbox}%
\framebox[0.9\textwidth]{\usebox{0}}%
\par\medskip
\ignorespacesafterend}
\newcommand{\bb}{\begin{mymathbox}}
\newcommand{\eb}{\end{mymathbox}}

\newcommand{\munit}{\mbox{\boldmath $1\!\!1$}}
\newcommand{\bs}{\boldsymbol}

\newcommand{\pslash}{\slashed{p}}

\newcommand{\be}{\begin{equation}}
\newcommand{\ee}{\end{equation}}
\newcommand{\ba}{\begin{eqnarray}}
\newcommand{\ea}{\end{eqnarray}}

\newcommand{\uu}{u({\bf p},s)}

\newcommand{\ubar}{\overline{u}({\bf p},s)}

\newcommand{\nk}{{\bf      k}}
\newcommand{\np}{{\bf      p}}
\newcommand{\nq}{{\bf      q}}
\newcommand{\nr}{{\bf      r}}

\newcommand{\nz}{{\bf      z}}

\newcommand{\npsi}{{\bf \npsi}}

\newcommand{\Psib}{\overline{\Psi}}
\newcommand{\psib}{\overline{\psi}}

\newcommand{\de}{\text{d}}

\newcommand{\non}{\nonumber}

\newcommand{\bma}{\begin{pmatrix}}
\newcommand{\ema}{\end{pmatrix}}

\newcommand{\Cab}{{\cos\theta_c}}

\begin{document}

\title{Lectures on lepton-nucleus quasielastic scattering with relativistic models}

\author{Raúl~Gonz\'alez-Jim\'enez\footnote{raugj@us.es}}
\affiliation{Departamento de F\'isica~At\'omica,~Molecular~y~Nuclear, Facultad de F\'isica, Universidad de Sevilla, Sevilla 41012, Spain}


\begin{abstract}
In these notes, basic concepts and necessary formulas for the modeling of elastic lepton-nucleon and quasielastic lepton-nucleus scattering using relativistic models are discussed. Certain theoretical developments are presented in meticulous detail, particularly those rarely found in papers or standard textbooks. In order to highlight the discrepancies between models and evaluate the impact of various nuclear effects, theoretical predictions are compared with inclusive electron scattering data.

These notes are designed for Master's or PhD students embarking on their research in the field of neutrino-nucleus interactions at intermediate energies (ranging from several hundred MeV to a few GeV).
They are intended to serve as a supplemental guide, never as a substitute for traditional textbooks on Scattering Theory and Quantum Field Theory.
Currently, these notes are included in the course `Relativistic Quantum Theory: Nuclear Processes', which is integrated into both the \href{https://master.us.es/fisicanuclear/index.php/}{Interuniversity Master in Nuclear Physics} and the \href{https://www.emm-nucphys.eu/en}{Erasmus Mundus Joint Master Degree in Nuclear Physics}, with the University of Seville as one of the partner institutions. Previously, this material was part of the course `Weak Interactions', within the same Master's programs.
\end{abstract}
\maketitle

\newpage 

\tableofcontents

\newpage 

\section{Introduction}

These notes are about the modeling of relativistic scattering of leptons by nucleons and nuclei.
For a more comprehensive and detailed discussion, the standard textbooks on the relativistic scattering of particles are recommended. Here, we closely follow~\cite{GreinerQED} for the conventions and derivation of the mathematical expressions of the lepton scattering cross section, but~\cite{HalzenMartins, MandlShaw, Foundations17, Bjorken} are also recommended for this and other aspects of the problem.

The relativistic scattering of leptons is a broad subject. Here, we focus on the elastic scattering of leptons by nucleons, accounting for their internal structure and working within a fully relativistic framework. The lepton-nucleon reaction serves as the foundation for the more complex lepton-nucleus {\bf quasielastic (QE)} scattering, which is the ultimate goal of these notes.

In the QE scattering, one assumes that the lepton couples a single nucleon which is bound in a nucleus, this nucleon is knocked out and the rest of nucleons, in first approximation, are considered as simple spectators. As in the elastic lepton-nucleon scattering, in the QE process, the struck nucleon is not excited and no other particles are created.

Describing the QE process requires to model the target nucleus, for that we will use relatively simple approaches, which despite their simplicity, are today widely used in the field of neutrino-nucleus interactions (some recent and comprehensive review articles are \cite{Alvarez-Ruso18,Alvarez-Ruso25}). 
In these notes, we discuss the differences between each other and how these affect the cross section results. The cross-section model predictions are compared to electron-scattering experimental data.

\section{Cross section}

The cross section informs us about the probability that for a given configuration of (two) particles in the initial state we will find a certain configuration of particles in the final state. For example, an electron beam with a given energy, $\varepsilon_i$, is made to hit a hydrogen target: the differential cross section with respect to the variables of the final (scattered) electron, denoted as $\frac{d^3\sigma(\varepsilon_i)}{d\varepsilon_fd\Omega_f}$, tells us how likely it is to find the scattered electron in a given solid angle $d\Omega_f$ with a certain energy $\varepsilon_f$; if we are not interested in the energy of the final lepton, then it is integrated and one gets $\frac{d^2\sigma(\varepsilon_i)}{d\Omega_f}$; and so on, the same reasoning applies for different variables.

Technically, the cross section is defined as the transition probability per particle and per unit of time of going from an initial-state configuration $i$ to a final-state $f$, divided by the incoming current of particles:
\ba
    d\sigma(\varepsilon_i) = \frac{|S_{fi}|^2}{\Phi_{inc}T} dN_f\,.\label{dsig_1}
\ea
In what follows, the different elements in eq.~\ref{dsig_1} are briefly described:

+ $T$ is a normalization time: the elapsed time for the reaction to occur.

+ $|S_{fi}|^2$ expresses the probability density of going from the initial state $i$ to the final state $f$. 

+ $dN_f$ is the number of final states, so that $|S_{fi}|^2dN_f$ is a probability. 

+ $\Phi_{inc}$ is the incident flux, given by the relative velocity of incident particles per unit volume: 
\ba
    \Phi_{inc}=\dfrac{\left| \bs v_i - \bs{\cal V}\right|}{V} = \dfrac{\left|\dfrac{\nk_i}{\varepsilon_i} - \dfrac{\np}{E}\right|}{V}\,.\label{flux1}
\ea
$V$ is the volume where the process occurs, $\bs v_i$ and $\bs{\cal V}$ are the velocities of initial lepton and hadronic target respectively. The notation for describing the particle 4-vectors is shown in Fig.~\ref{fig:elastic} and Appendix~\ref{app:Conventions}.  
Eq.~\ref{flux1} is valid only for collinear particles. It can be further simplified if we work in a reference frame where the target is at rest ($\np=0$) and the beam is made of ultrarelativistic particles ($k_i\approx \varepsilon_i$), in that case: 
\ba\Phi_{inc}=1/V\,.\ea
If you are now puzzled by what happened with the units in the expression above (a velocity disappear from the numerator?), don't worry, it is {\it natural}... (see eq.~\ref{hbc} in Appendix~\ref{app:Conventions}).

A Lorentz covariant expression of the cross section (i.e. valid in any reference frame, not only for collinear particles) is possible if one uses a slightly different definition of the flux:
\ba
    \Phi_{inc} = \dfrac{1}{V}\dfrac{\sqrt{(K_i\cdot P)^2 - m_i^2 M_i^2} }{\varepsilon_i E}\,,
\ea
where $m_i$ and $M_i$ are the masses of the initial lepton and nucleon.
(See discussion starting at pg. 111 in~\cite{GreinerQED} for more details.)

\section{Elastic lepton-nucleon scattering}\label{sec:elastic}

\begin{figure}[htbp]
\centering  
\includegraphics[width=.5\textwidth,angle=0]{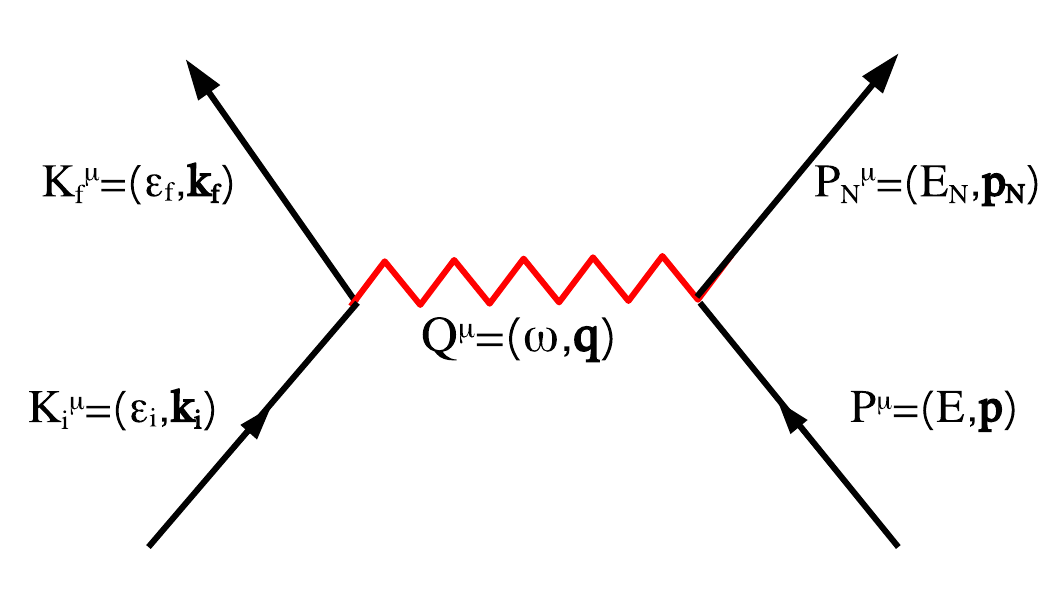}
\caption{One-boson exchange diagram for elastic lepton-nucleon scattering. $K_i^\mu=(\varepsilon_i,\nk_i)$, $K_f^\mu=(\varepsilon_f,\nk_f)$, $P^\mu=(E,\np)$, $P_N^\mu=(E_N,\np_N)$ and $Q^\mu=(\omega,\nq)$ are the 4-momenta of the initial and final lepton, initial and final nucleon, and virtual boson, respectively.}\label{fig:elastic}
\end{figure}

The elastic electron-nucleon scattering in the Born approximation, i.e., only one boson exchanged, is sketched in Fig.~\ref{fig:elastic}. Using the Feynman rules given in Appendix~\ref{Feyn-rules}, the matrix element reads:    
\ba
    S_{fi} &=& i\int{d^4X}\, i\int{d^4Y}\, \int{\frac{\de^4Q}{(2\pi)^4}}\,\non\\
      &\times& e\Psib_{e'}(X)\gamma^\mu\Psi_{e}(X) \,\,\,\,\,\,\,\,\,\,\,\,\,\,\,\,\,\,\,\,\,\,\,\,\,\,\,\,\,\,\,\,\,\,\,\, \text{(lepton current)} \non\\
      &\times& \frac{-i g_{\mu\nu}}{Q^2}e^{iQ\cdot(X-Y)} \,\,\,\,\,\,\,\,\,\,\,\,\,\,\,\,\,\,\,\,\,\,\,\,\,\,\,\,\,\,\,\,\,\,\,\,\,\, \text{(photon propagator)} \\ 
      &\times& -e\Psib_{N'}(Y)\Gamma^\mu(P_N-P)\Psi_{N}(Y) \,\,\,\,\,\,\,\,\, \text{(hadron current)}\non
\ea
Since all external particles are free, we described them as plane waves. Then, all the integrals can be trivially performed (see  the definition of the fermion wave functions in Appendix~\ref{Feyn-rules} and eq.~\ref{Delta}). 
This results in the following expression: 
\ba
    S_{fi} &=& -i \sqrt{\frac{m_i}{\varepsilon_i V}} \sqrt{\frac{m_f}{\varepsilon_f V}} 
    \sqrt{\frac{M_i}{EV}}\sqrt{\frac{M_f}{E_NV}} \non\\
    &\times& (2\pi)^4\delta^4(K_f+P_N-K_i-P)\frac{ej_\mu\, eJ^\mu}{Q^2}
\ea
with $M_{i,f}$ and $m_{i,f}$ the initial and final nucleon and electron masses. 
We have introduced the leptonic and hadronic currents defined as:
\ba
    j^\mu &\equiv& \bar{u}(\nk_f,s_f)\gamma^\mu u(\nk_i,s_i)\,,\\
    J^\mu &\equiv& \bar{u}(\np_N,s_N)\Gamma^\mu(Q^2) u(\np,s)\,.
\ea
$\gamma^\mu$ are the Dirac matrices and $\Gamma^\mu$ is a generic function that is described later. They are usually called the lepton- and hadron-current operators. 

Assume that we want to describe the {\bf elastic scattering of an electron by a proton}, typically an electron beam hitting a hydrogen target. Assume that the initial electron is not polarized, so we do not know its third spin component. Consider the target is unpolarized too. Finally, consider that we do not measure the spin of the final electron or final proton. In this case, to compute the cross section of our scattering process we need to average over the possible initial states and sum over the possible final states~\footnote{Note that these summations are done at the cross section level, not at the amplitude level, so these are {\it incoherent sums} (no interferences). This is so because final states made of different spin configurations are {\it different final states}, i.e., one could distinguish them, in this case, by measuring the spin.}. 

Hence, for the cross section we need $\overline{\sum}|S_{fi}|^2 = \overline{\sum}(S_{fi})^*S_{fi}$, where $\overline{\sum}$ implies an average over all possible initial states and sum over all possible final states. In our case, this is an average and sum over the two helicity states of the electron and proton~\footnote{To insist on this, we can imagine an experimental set up such that the incoming electron beam is polarized, i.e., initial electrons are in a defined helicity state. In that case, we do not average over the two possible spin states of the electron because we know in which state it is. 
Notice that for initial neutrinos (antineutrinos) one should not perform the average on the spins since they are left-handed (right-handed) particles.
}. 
One gets:
\ba
    \overline{\sum}|S_{fi}|^2 = \dfrac{m_im_f}{\varepsilon_f\varepsilon_i}\dfrac{M_iM_f}{EE_N} \dfrac{e^4}{Q^4}\,
     \dfrac{T}{V} (2\pi)^4\delta^4(K_f+P_N-K_i-P) L_{\mu\nu}H^{\mu\nu}\,,\label{Sfi}
\ea
with the lepton and hadron tensors defined by
\ba
    L_{\mu\nu}\equiv\frac{1}{2}\sum_{s_i}\sum_{s_f}(j_\mu)^*j_\nu\,,\,\,\,\,\, 
    H^{\mu\nu}\equiv\frac{1}{2}\sum_{s}\sum_{s_N}(J^\mu)^*J^\nu.
\ea

Doing the algebra, one gets the following expressions:
\ba
L_{\mu\nu} &=& \frac{1}{2m_im_f} \left[ K_{i,\mu} K_{f,\nu} + K_{i,\nu} K_{f,\mu} - g_{\mu\nu}(K_i\cdot K_f+m_im_f) \right]\,,\label{Lmn}\\
H^{\mu\nu} &=& \frac{1}{2M_iM_f} \text{Tr}\left[\frac{(\displaystyle{\not}P_i+M_i)}{2} 
\overline{\Gamma}^{\mu}_{EM} 
\frac{(\displaystyle{\not}P_f+M_f)}{2}\Gamma^{\nu}_{EM}\right]\,.
\label{HmnT}
\ea
(See~\cite{GreinerQED} or \cite{Bjorken} for details on the calculation of these quantities.) The subindex $EM$ in the hadron current operator means {\it electromagnetic interaction}.

Most of the time, we work with leptons with very small masses (compared to their momenta), such as neutrinos or ultrarelativistic electrons. Thus, for computational reasons, it is convenient to eliminate the lepton masses from the expressions (actually, from the denominators), so they can be implemented into our codes for numerical calculations. Therefore, we redefine the lepton tensor as:
\ba
    L'_{\mu\nu} &\equiv& m_im_f L_{\mu\nu} = \frac{1}{2} \left[ K_{i,\mu} K_{f,\nu} + K_{i,\nu} K_{f,\mu} - g_{\mu\nu}(K_i\cdot K_f+m_im_f) \right]\,,
\ea
where the masses that appeared in eq.~\ref{Sfi} have been absorbed (and canceled) in the lepton tensor. 

Nucleons are not point-like elementary particles. The function $\Gamma^{\mu}_{EM}$ accounts for their internal structure (see Appendix~\ref{app:structure}). One can find different forms of this function in the literature. Here, we use the so-called CC1 operator: 
\ba
\left. \Gamma^{\mu}_{EM}\right|^{p,n}_{CC1}  =
(F_1^{p,n}+F_2^{p,n})\gamma^{\mu}-\frac{F_2^{p,n}}{2M}(P+P_N)^\mu \,,\label{CC1}
\ea
the superscript `$p,n$' denotes proton (p) or neutron (n). 
With this current operator, one obtains the following expression for the hadron tensor (we omit the index $p,n$ in what follows to simplify the notation, but remember that the cross section corresponds to elastic electron-proton or electron-neutron scattering):
\ba
 H^{\mu\nu} &=& \frac{1}{2M^2} \Bigg\lbrace (F_1+F_2)^2 \left(P^{\mu}P_N^{\nu}+P^{\nu}P_N^{\mu} 
    +\left(M^2 - P\cdot P_N \right) g^{\mu\nu}\right) \non \\
   &+& \left[\left(\frac{F_2}{2M}\right)^2\left(P\cdot P_N + M^2\right) - F_2(F_1+F_2)\right]
   (P+P_N)^{\mu}(P+P_N)^{\nu} \Bigg\rbrace\,. \label{SMN}
\ea
From now on, we neglect the subindex $i,f$ in the lepton and nucleon mass~\footnote{In the case of elastic or QE scattering induced by neutral currents, as the EM interaction, the initial and final particles are the same, hence, they have the same masses. In the case of charged-current interaction, this is not the case; however, given the high energies involved in the processes studied here, one can safely neglect the difference between proton and neutron masses.}.
The analogous expression using an alternative, so called, CC2 operator (see Appendix~\ref{app:structure}) gets more involved; however, for on-shell nucleons, one gets exactly the same cross section  with the CC1 and CC2 operators.\\

All together, the six-differential cross section reads~\footnote{The number of final states within a range of momenta $d\np_Nd\nk_f$ is (see page 85 in~\cite{GreinerQED} for more details):
\ba 
dN_f=\frac{V}{(2\pi)^3}d\np_N\,\frac{V}{(2\pi)^3}d\nk_f.\label{numberstates}
\ea
}:
\ba
    d^6\sigma(\varepsilon_i) = K  (2\pi)^4\delta^4(K_f+P_N-K_i-P) L'_{\mu\nu}H^{\mu\nu}\dfrac{d\nk_f}{(2\pi)^3} \dfrac{d\np_N}{(2\pi)^3}\label{d6sig}
\ea
with 
\ba
    K &\equiv& \dfrac{1}{V\Phi_{inc}} \dfrac{1}{\varepsilon_f\varepsilon_i}\dfrac{M^2}{EE_N} \dfrac{e^4}{Q^4}\non\\
      &=& \dfrac{1}{\sqrt{(K_i\cdot P)^2-m^2M^2}}\dfrac{M^2}{\varepsilon_fE_N} \left(\frac{4\pi\alpha}{Q^2}\right)^2\,.
\ea
We have introduced the fine-structure constant $\alpha=\dfrac{e^2}{4\pi}=\dfrac{1}{137.036}$. 

Conservation of 4-momentum is ensured by the Dirac delta (after integration over 4 of the remaining variables).
One is free to choose the independent variables that will describe the process. 
The decision on which variable to integrate depends on the observables we want to compare with. 
For example, if an experiment measures the scattered lepton, then one integrates over the final nucleon variables, which results:
\ba
  \frac{d^3\sigma(\varepsilon_i)}{d\nk_f} = \frac{K}{(2\pi)^2} \delta(\varepsilon_f + E_N-\varepsilon_i-E) L'_{\mu\nu}H^{\mu\nu}\,.\label{d3sigl}
\ea
If what was measured is the final nucleon (as in the case of a neutrino-induced neutral-current scattering), then one integrates over the final lepton variables, giving:
\ba
    \frac{d^3\sigma(\varepsilon_i)}{d\np_N} = \frac{K}{(2\pi)^2}  \delta(\varepsilon_f + E_N-\varepsilon_i-E) L'_{\mu\nu}H^{\mu\nu}\,.\label{d3sigN}
\ea

The energy-conservation Dirac delta in eqs.~\ref{d3sigl} and \ref{d3sigN} is used to analytically integrate over one of the remaining variables; one chooses it, it depends on the goal.
As an example, imagine an experiment in which the scattered lepton is detected at a given solid angle. Hence, one is interested in the double-differential cross section $\frac{d^2\sigma(\varepsilon_i)}{d\Omega_f}$, i.e., we need to integrate eq.~\ref{d3sigl} over $k_f$.

\noindent\rule{4cm}{0.4pt}

{\it Let's do it, step by step.}

First, remember that in spherical coordinates $d\np = p^2dp d\Omega$, where $d\Omega=d\cos\theta d\phi$ is a differential of solid angle and $\theta$ and $\phi$ are the polar and azimuthal angles, respectively.

We chose to integrate over $k_f$ or, equivalently, over $\varepsilon_f$ (using 
$\varepsilon_fd\varepsilon_f=k_fdk_f$), so we need to write the energy-conservation Dirac delta as a function of the independent variables: $\nk_i=k_i(\sin\theta_i\cos\phi_i, \sin\theta_i\sin\phi_i,\cos\theta_i)$ and $\nk_f=k_f(\sin\theta_f\cos\phi_f, \sin\theta_f\sin\phi_f,\cos\theta_f)$.
We will work in the lab frame, where the target nucleon is at rest, $\np=0$, and $\nk_i$ defines the $\hat\nz$ axis, $\nk_i=(0,0,k_i)$. In this case:
\ba
  \delta[f(\varepsilon_f)] = \delta\left(\varepsilon_f - \varepsilon_i + \sqrt{k_i^2 + k_f^2 - 2k_i k_f\cos\theta_f + M^2} - M\right)\,,\label{Kinelastic1}
\ea
where we have used 
\ba
  E_N &=& \sqrt{p_N^2+ M^2} = \sqrt{q^2 + M^2} = \sqrt{k_i^2 + k_f^2 - 2k_i k_f\cos\theta_f + M^2}\,.
\ea
In the ultrarelativistic limit for the electrons ($\varepsilon/k\rightarrow1$), one gets:
\ba
  f(\varepsilon_f) = \varepsilon_f - \varepsilon_i + \sqrt{\varepsilon_i^2 + \varepsilon_f^2 - 2\varepsilon_i\varepsilon_f\cos\theta_f + M^2} - M\,.\label{Kinelastic}
\ea

Hence,
\ba
  f_{rec}\equiv\left|\frac{\partial f(\varepsilon_f)}{\partial\varepsilon_f}\right| = \left|1 + \frac{\varepsilon_f -\varepsilon_i\cos\theta_f}{E_N}\right| \,.\label{frec0}
\ea
$f_{rec}$ is the so-called {\it recoil} factor, called like that because of the following. If the mass of the target was very large compared to the lepton energies, then the hadron system would receive very little kinetic energy ($E_N\approx M$), or in other words, there would be very little recoil. In that case, $(\varepsilon_f-\varepsilon_i\cos\theta_f)/E_N<<1$ so $f_{rec}\rightarrow1$. In contrast, if the lepton energies are comparable or larger than the nucleon mass, it may receive considerable amount of kinetic energy, so the recoil factor could be very different from 1.

Finally, using eq.~\ref{delta-prop}, the Dirac delta reads:
\ba
  \delta[f(\varepsilon_f)] = \dfrac{\delta(\varepsilon_f-\varepsilon_f^0)}{\left|\frac{\partial f(\varepsilon_f)}{\partial\varepsilon_f}\right|_{\varepsilon_f=\varepsilon_f^0}}\,.
\ea  
Where $\varepsilon_f^0$ is obtained by solving the equation $f(\varepsilon_f)=0$ for $\varepsilon_f$:
\ba
\varepsilon_f^0 = \frac{\varepsilon_i}{1 + \frac{\varepsilon_i}{M}(1-\cos\theta_f)}\,.
\ea
If we do not consider the ultrarelativistic limit for the leptons, the expressions are more involved, see Appendix~\ref{full-kinematics}.

Summarizing, after integrating the Dirac delta, the double-differential cross section reads:
\ba
    \frac{d^2\sigma(\varepsilon_i)}{d\Omega_f} = \frac{K}{(2\pi)^2} \frac{k_f\varepsilon_f}{f_{rec}} L'_{\mu\nu}H^{\mu\nu}\,.\label{d2sig}
\ea

\noindent\rule{4cm}{0.4pt}

\begin{figure}[htbp]
\centering  
\includegraphics[width=.5\textwidth,angle=0]{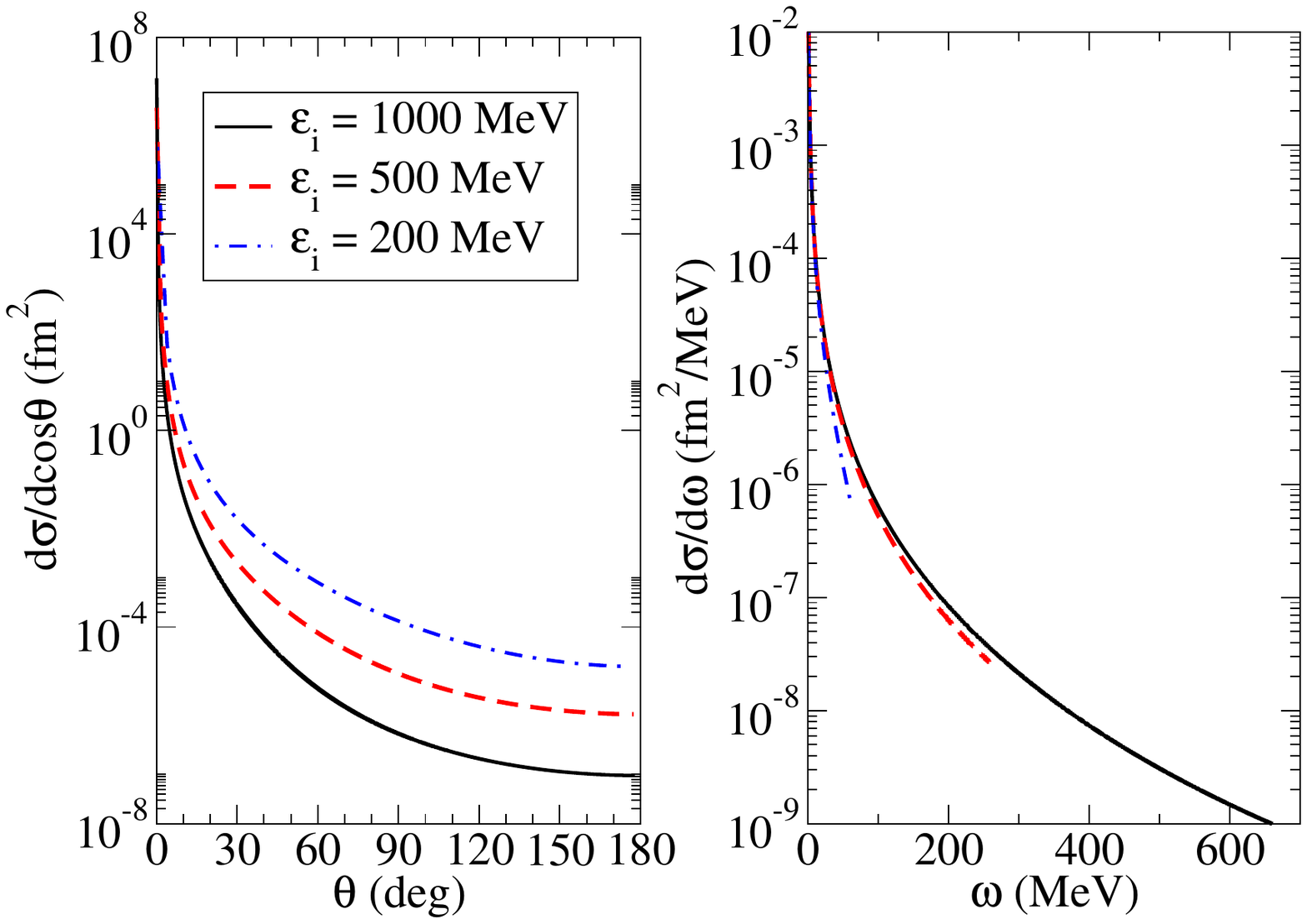}
\vspace{-0.5cm}
\caption{Elastic electron-proton single differential cross section. The results correspond to three values of the incoming electron energy: $\varepsilon_i=$ 200, 500 and 1000 MeV. The left (right) panel is the cross section as a function of the cosine of the scattering angle (energy transfer).}\label{fig:elastic_ep}
\end{figure}

As an example, in Fig.~\ref{fig:elastic_ep} we show the elastic electron-proton scattering cross section for several values of the energy of the beam in terms of the electron scattering angle (left panel) and the energy transfer (right panel). One observes that the cross section decreases rapidly as the scattering angle and energy transfer increase.

The double-differential cross section is often written in the equivalent form:
\ba
    \frac{d^2\sigma(\varepsilon_i)}{d\Omega_f} = \sigma_{Mott} \left(\frac{\varepsilon_f}{\varepsilon_i}\right)\left(v_LR_L+ v_TR_T\right)\,.\label{dsig_LT}
\ea
This is called the Rosenbluth decomposition. Details and explicit expressions of the Mott cross section ($\sigma_{Mott}$), the leptonic factors ($v_{L,T}$), the hadronic response functions $R_{L,T}$, etc. can be found e.g. in Refs.~\cite{Foundations17} and \cite{GreinerQED}.

As final remarks:
\begin{itemize}
 \item It is important to understand that by choosing an appropriate reference frame, the cross section does not depend on one of the azimuthal angles, $\phi_f$ or $\phi_N$. For example, if one chooses the lepton 3-vectors to define the $xz$-plane, the cross section does not depend on $\phi_f$. 
 \item The derivation of the cross section formula and kinematics performed here is valid for any 2 to 2 scattering process. The pieces that depend on the particular reaction are the particle description (e.g. masses) and the matrix element, $S_{fi}$, which includes Feynman diagrams, coupling constants, particle propagator, etc.
\end{itemize}

\section{Quasielastic scattering. The impulse approximation}

\begin{figure}[htbp]
\centering  
\includegraphics[width=.5\textwidth,angle=0]{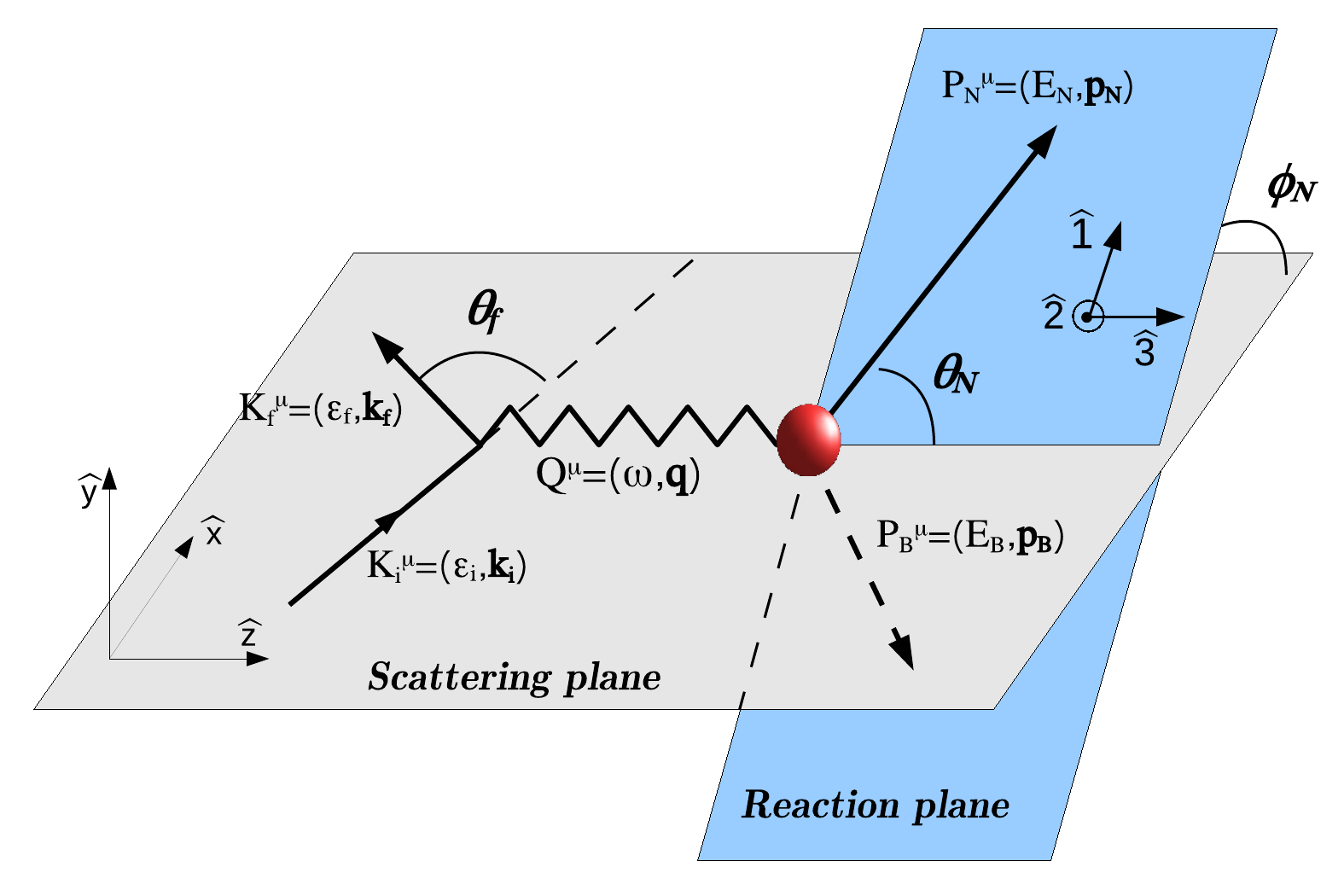}
\caption{Sketch of a quasielastic lepton-nucleon scattering in a reference frame in which $\nq$ is along $\hat\nz$ and the leptons define the xz-plane.
}\label{fig:QE_planos}
\end{figure}

In quasielastic (QE) scattering, sketched in Fig.~\ref{fig:QE_planos}, the incident lepton ($K_i$) interacts with a target nucleus ($P_A$). The final state consists of the scattered lepton ($K_f$), a knockout nucleon ($P_N$) and a residual system ($P_B$). 
As we did when we modeled the lepton-nucleon elastic interaction, we work in the Born approximation, i.e., we consider that only one boson is exchanged ($Q$). 

\begin{figure}[htbp]
\centering
\includegraphics[width=.7\textwidth,angle=0]{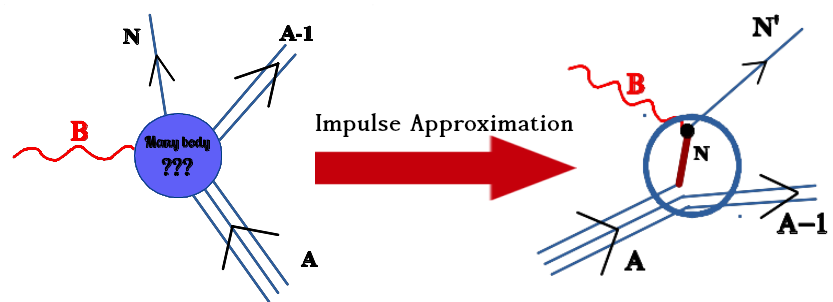}
\caption{The Impulse Approximation for QE scattering.}\label{fig:IA}
\end{figure}

The modeling of the hadron current is, in this case, a complex many-body problem: 
\ba
  J^\mu = \langle N, A-1|{\cal O}^\mu_{\text{many-body}}|A\rangle\,.
\ea
To deal with it, we work in the impulse approximation (IA). It consists in considering that the boson couples only a bound nucleon of the target nucleus, the rest of nucleons are simple spectators (see Fig.~\ref{fig:IA}). The IA allows us to move from a many-body to a one-body problem, since now, the hadron current reads:
\ba
  J^\mu = \langle N'|{\cal O}^\mu_{\text{one-body}}|N\rangle\,,
\ea
and the QE process is described as the incoherent sum of ${\cal N}$ one-body processes, ${\cal N}$ being the number of active nucleons in the process (protons, neutrons or both), and in ${\cal O}^\mu_\text{one-body}$ one uses the same  operator as in the elastic lepton-nucleon scattering, i.e., eq.~\ref{CC1} for the EM interaction.

However, one should not forget that the interaction occurs inside the nucleus; therefore, the initial nucleon is bound and not at rest, and the final one may suffer final-state interactions on its way out of the nucleus.  
In summary, to describe QE scattering we need a nuclear model. In what follows, we list some (from simpler to more complex): 
\begin{itemize}
 \item {\bf The global relativistic Fermi gas (gRFG)}: Gas of non-interacting fermions in the thermodynamic limit, i.e. an infinitely large volume and number of particles, but constant density $\rho=N/V$. 
 In this approach, the Fermi momentum (or Fermi energy) is a constant, for a given temperature, and the only parameter in the model. Nucleons are described as on-shell relativistic plane waves. The model accounts for Fermi motion and Pauli's exclusion principle. Binding energy must be incorporated ad hoc.
 \item {\bf The local relativistic Fermi gas (lRFG)}: The starting point is a gas of non-interacting fermions in the thermodynamic limit at zero temperature. 
 Then, the constant density is replaced by a `realistic' nuclear density $\rho(r)$, being $r$ the nuclear radius. 
 As a result, the Fermi momentum, which is a constant parameter in the gRFG, is now a function of $r$. Nucleons are described as on-shell relativistic plane waves. The model accounts for Fermi motion. Binding energy and Pauli's exclusion principle must be incorporated ad hoc.
 \item {\bf Plane-wave impulse approximation (PWIA) with a realistic spectral function}: Realistic description of the nuclear initial state, which is factorized from the reaction vertex. The initial and final nucleon are described as plane waves. It accounts for Fermi motion, binding energy and other nuclear effects all related to the initial sate. Pauli's exclusion principle must be added ad hoc.
 \item {\bf Relativistic distorted-wave impulse approximation (RDWIA)}: Initial- and final-state nucleons are described by wavefunctions which are solution of the wave equation in the presence of relativistic mean-field potentials. The model accounts for Fermi motion, binding energy, Pauli's exclusion principle, other nuclear effects in the initial state and final-state interactions between the struck nucleon and the residual system. 
\end{itemize}

\section{Quasielastic scattering within the relativistic global Fermi gas model}\label{sec:gRFG}

The relativistic Fermi gas (RFG) model, also known as {\it global} RFG to differentiate it from its {\it local} brother (Section~\ref{sec:lRFG}), is probably the simplest relativistic nuclear model. It is suited for the ideal case of {\it infinite nuclear matter}. Though, it provides rough estimates of some basic nuclear properties and it is fully relativistic.
The gRFG consists in a gas of non-interacting identical fermions in the thermodynamic limit. Nucleons are described by Dirac on-shell plane waves. The quantum numbers are the spin projection and momentum of the nucleons. The energy spectrum of nucleons is a continuum (the Fermi sea).

The number of particles for a given energy level is given by the Fermi-Dirac distribution function~\footnote{For more information about the Fermi-Dirac distribution, a detail derivation and deeper discussion, see any book about Statistical Mechanics. But one can start with the Wiki~\cite{Wiki-FermiDirac}.}:
\ba
  f(p,T) = \frac{1}{1 + \exp\left(\frac{E - E_F}{k_BT}\right)} = \frac{1}{1 + \exp\left(\frac{\sqrt{p^2+M^2} - \sqrt{p_F^2+M^2}}{k_BT}\right)}\,,
\ea
where $p_F$ is the Fermi momentum. the only free parameter of the model. $k_B=8.62\times10^{-11}$ MeV/K is the Boltzmann constant and $T$ the temperature. 
The total number of particles in the system results from the sum over all quantum states:
\ba
  {\cal N} = \sum_\alpha f(p,T)\,.
\ea
We need to convert the sum over discrete states $\sum_\alpha$ into an integral over the continuum momentum states and a sum over spin. This is done by the following replacement (see footnote~\ref{numberstates}): 
\ba
\sum_\alpha\longrightarrow\frac{V}{(2\pi)^3}\int d\np\,\sum_s\,,\label{continuouslimit}
\ea
where $V$ is the volume, $s$ is the spin projection and $\np$ is the momentum. 
Hence, the number of particles in the gRFG model is:
\ba
  {\cal N} = \frac{V}{(2\pi)^3}\int d\np\,\sum_s f(p,T) = 2\frac{V}{(2\pi)^3}\int d\np\,f(p,T)\,.
\ea

We define the {\bf momentum distribution}, $n(p)$, as the probability density function of finding a nucleon in the nucleus with a given momentum $p$. Therefore, by definition:
\ba
  \int d\np\, n(p) = {\cal N}\,.
\ea
Therefore, in the gRFG one gets:
\ba
 n_{gRFG}(p,T) = 2\frac{V}{(2\pi)^3}f(p,T)\,.
\ea 

Integrating ($\int d\np$) on the lhs and rhs the equation above, we obtain an explicit expression for the volume, which is a function of $T$:
\ba
  V(T) = (2\pi)^3\frac{{\cal N}}{2{\cal I}(T)}\,,
\ea
where 
\ba
  {\cal I}(T) = \int d\np f(p,T)\label{I(T)}
\ea
is a finite number which could be computed numerically or, in some cases, analytically.
Thus, the momentum distributions is:
\ba
 n_{gRFG}(p,T) =\frac{{\cal N}}{{\cal I}(T)} f(p,T)\,.\label{nRFG}
\ea 

At zero temperature ($T=0$ K), $f(p,T=0)$ is a step function, in other words, all levels up to the Fermi energy are fully occupied and empty beyond that. In this case, the integral of eq.~\ref{I(T)} can be trivially done, giving ${\cal I}(T=0)=\frac{4}{3}\pi p_F^3$. Hence, the volume reads:
\ba
  V = (2\pi)^3\frac{{\cal N}}{2\frac{4}{3}\pi p_F^3}\,;
\ea
consequently, the momentum distribution is given by:
\ba
 n_{gRFG}(p) = \frac{{\cal N}}{\frac{4}{3}\pi p_F^3}\Theta(p_F-p). 
\ea

The {\bf nuclear density} is defined as the number of particles per unit volume and, in general, is a function of the position $\nr$:
\ba
\rho(\nr)\equiv\frac{{\cal N}}{V} = \sum_\alpha\Psi_\alpha^\dagger(\nr)\Psi_\alpha(\nr)\,,
\ea
where $\Psi_\alpha(\nr)$ is the wavefunction of a bound nucleon with quantum number(s) $\alpha$.

In the gRFG, for a given temperature, the nuclear density is a constant: 
\ba
  \rho(T) = \frac{{\cal N}}{V(T)} = 2\frac{{\cal I}(T)}{(2\pi)^3}\,.
\ea
When $T=0$ K one gets
\ba
  \rho = \frac{{2\frac{4}{3}\pi p_F^3}}{(2\pi)^3}\,. \label{rho_pF}
\ea

\begin{figure}[htbp]
\centering  
\includegraphics[width=.47\textwidth,angle=0]{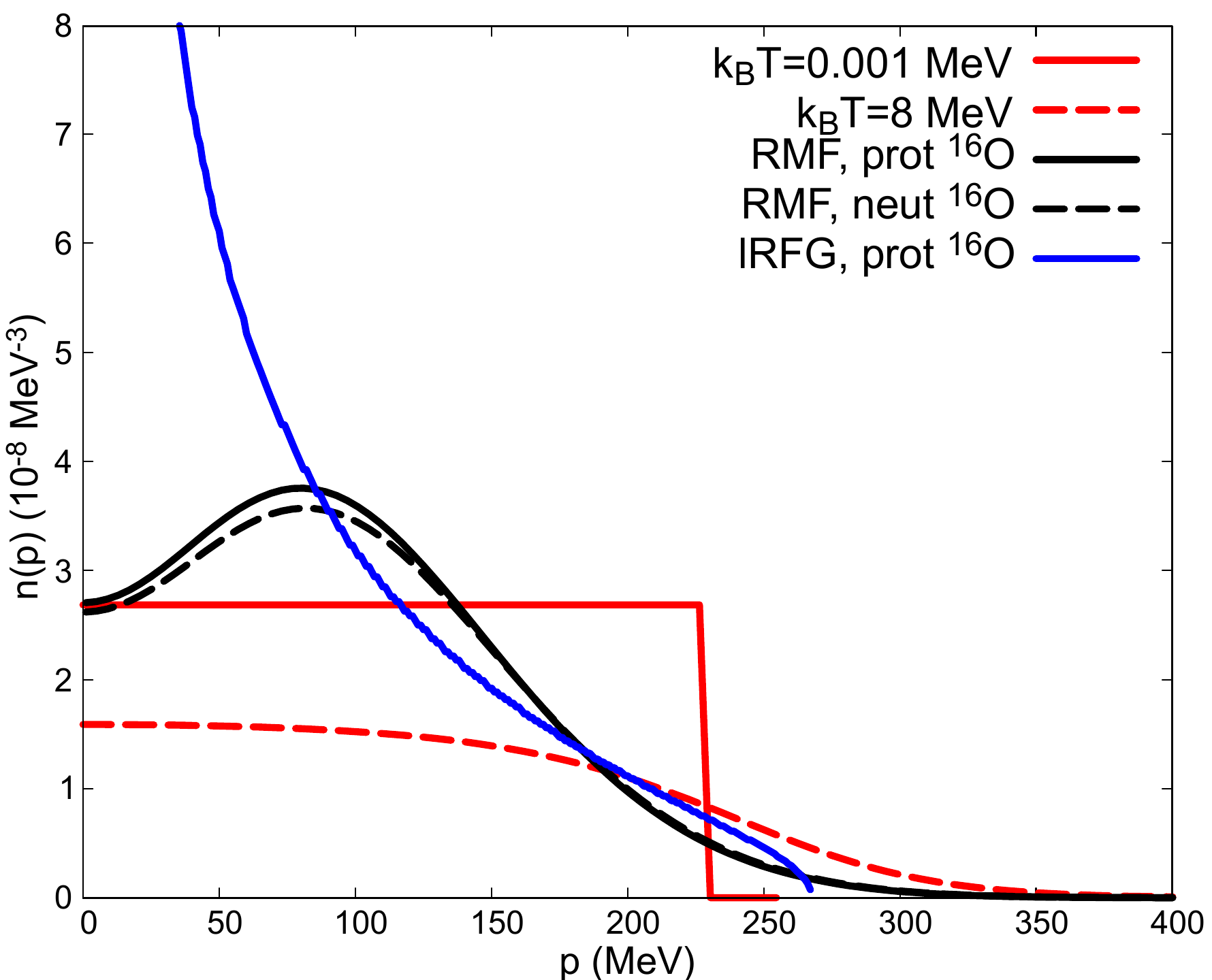}
\includegraphics[width=.47\textwidth,angle=0]{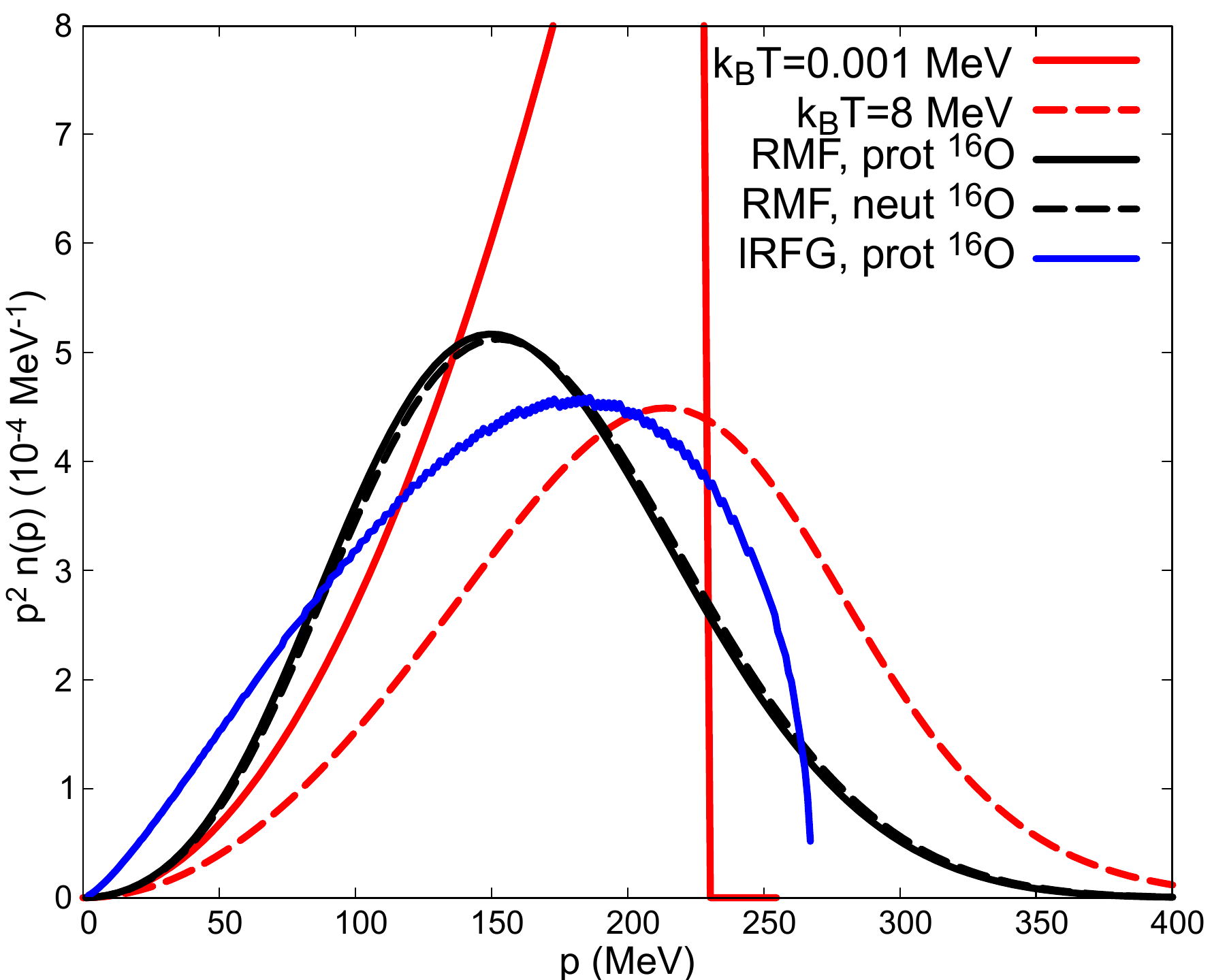}
\caption{Momentum distributions $n(p)$ of nucleons derived with the gRFG model of Section~\ref{sec:gRFG} for $k_B T\approx0$ and 8 MeV, and with the lRFG model from Section~\ref{sec:lRFG}. The proton and neutron momentum distributions from a relativistic mean-field (RMF) model are also shown for comparison. All $n(p)$ are normalized to 1 ($\int d\np\, n(p) = 1$). In the right panel the phase-space factor $p^2$ has been included.}\label{fig:mom-dist}
\end{figure}

In Fig.~\ref{fig:mom-dist} we show the gRFG momentum distributions for the cases of $k_BT=8$ MeV (i.e., $T\approx10^{11}$ K, which is the typical temperature of the nucleus) and $T=0$ K. Also, we show the (more realistic) proton and neutron momentum distributions computed with a relativistic mean-field model. 
In the gRFG model, the parameter $p_F$ is taken from comparison with experiments. For example, it is related with the width of the QE peak in inclusive electron scattering cross sections. 
For $^{12}$C and $^{16}$O we use $p_F=228$ MeV, and for calcium 40 we use $p_F=241$ MeV (see more details in pg. 486 in~\cite{Foundations17}). 
For symmetric (N=Z) nuclei, it is common to use the same $p_F$ for neutrons and protons.

The QE cross section within the gRFG model is obtained by averaging~\footnote{Do you remember how to compute the average of a continuous function given a probability density function (PDF)?
Let $f$ be a function of $x$ and the continuous variable $y$, let $P(y)$ be the PDF of the variable $y$. Then, the average of $f$ over $y$ is given by:
\ba
 \langle f(x) \rangle = \frac{\int dy f(x,y) P(y)}{\int dy P(y)}\,, 
\ea
where the denominator is just the norm of the PDF.
} 
the elastic cross section off a free nucleon (eq.~\ref{d6sig}) over the 4-momentum of the initial nucleon.
We do it in what follows for the particular case of $T=0$ K:
\ba
    \frac{d^6\sigma(\varepsilon_i)}{d\nk_f d\np_N} &=& 
    \frac{{\cal N}}{4/3\pi p_F^3}\int dE\, \int d^3\np \,\, \Theta(p_F-p) \Theta(p_N-p_F)\,\, \delta(E-\sqrt{p^2+M^2}) \non\\
    &\times& \dfrac{K}{(2\pi)^2}  \delta^4(K_f+P_N-K_i-P) L'_{\mu\nu}H^{\mu\nu}\,.\label{d6sigRFG1}
\ea
Note that the factor $\dfrac{K}{(2\pi)^2}  \delta^4(K_f+P_N-K_i-P) L'_{\mu\nu}H^{\mu\nu}$ is just the elastic lepton-nucleon cross section (eq.~\ref{d6sig}), while the rest gives the average, i.e.: 
$\delta(E-\sqrt{p^2+M^2}) \frac{{\cal N}}{4/3\pi p_F^3}\Theta(p_F-p)$ is the probability density function (PDF) for the set of variables $P$,  the 4-vector describing the bound nucleon.

$\delta(E-\sqrt{p^2+M^2})$ is introduced to ensure that bound nucleons are on-shell particles, i.e., their energy and momentum are constrained by the relation $E=\sqrt{p^2+M^2}$ with $M$ the invariant nucleon mass. 
Note that in other approaches, beyond the RFG model, this is not so, and bound nucleons can be slightly off-shell.

The step function $\Theta(p_F-p)$
ensures that the cross section is zero if $p>p_F$, because there are no nucleons above the Fermi level.

Finally, $\Theta(p_N-p_F)$ is introduced to account for the Pauli's exclusion principle: it ensures that the cross section is zero if the knocked out nucleon (with momentum $\np_N$) is below the Fermi surface. This is because in the Fermi gas model, the Fermi sphere is completely filled with nucleons (all levels are occupied).

As always, one is free to choose the variables for integration. Four of the integrals must be done analytically using the 4-momentum conservation Dirac delta. 
For example, integration over the initial nucleon gives: 
\ba
    \frac{d^6\sigma(\varepsilon_i)}{d\nk_f d\np_N} &=& \frac{{\cal N}}{4/3\pi p_F^3} \Theta(p_N-p_F) \Theta(p_F-p)\non\\
    &\times& \dfrac{K}{(2\pi)^2} \delta(\varepsilon_f + E_N-\varepsilon_i-E) L'_{\mu\nu}H^{\mu\nu}\,,
\ea
where now, we have $\np=\nk_f+\np_N-\nk_i$. 
One additional analytical integration remains to be done, but in this case the variables inside the delta are not independent, so one needs to use the property in eq.~\ref{delta-prop}.

\noindent\rule{4cm}{0.4pt}

{\it Let's do an example:}

 We choose to integrate over $p_N$, so we need to write the delta as a function of the independent variables: $\nk_i$, $\nk_f$, and $\np_N$:
\ba
    \delta(\varepsilon_f + E_N-\varepsilon_i-E) &=& \delta(\sqrt{p_N^2+M^2}-\sqrt{|\np_N-\nq|^2+M^2} - \omega)\non\\
    &=& \delta(\sqrt{p_N^2+M^2}-\sqrt{p_N^2+q^2-2p_Nq\cos\theta_{p_Nq} + M^2} - \omega) \label{int_delta}
\ea
where we have used $\omega=\varepsilon_i-\varepsilon_f$ and $\nq=\nk_i-\nk_f$.
Now we use the property in eq.~\ref{delta-prop}:
\ba
    f(p_N) &=& \sqrt{p_N^2+M^2}-\sqrt{p_N^2+q^2-2p_Nq\cos\theta_{p_Nq} + M^2} - \omega\,,\\
    \left|\dfrac{\partial f(p_N)}{\partial p_N}\right| &=& \left| \frac{p_N}{E_N} - \frac{p_N-q\cos\theta_{p_Nq} }{E} \right|\,.
\ea
We need $p_N$ as a function of the independent variables, i.e., $p_N=p_N(\nk_i,\nk_f,\hat\np_N)$ with $\hat\np_N=(\sin\theta_N\cos\phi_N,\sin\theta_N\sin\phi_N,\cos\theta_N)$.
This is done by solving the equation $f(p_N) = 0$, that results:
\ba
    p_N = \dfrac{-AB \pm \sqrt{A^2 + (B^2-1)M^2} }{ (B^2-1) }\,,\label{pN_sol}
\ea
with 
\ba
    A \equiv \dfrac{Q^2}{2\omega}\,,\,\,\,\,
    B \equiv \dfrac{q\cos\theta_{p_Nq}}{\omega}\,.\non
\ea
There are two solutions for $p_N$, lets call them $p_N^+$ and $p_N^-$. A priori, both of them could have physical meaning, i.e.: $p_N$ is positive and solution of the original equation $f(p_N) = 0$. The interpretation is that for the given configuration $\nk_i$, $\nk_f$ and $\hat\np_N$ there are two values of the momentum of the final nucleon that fulfill energy-momentum conservation, therefore, if this momentum (or energy) of the final nucleon is not measured, the cross section will result from the sum over these two possibilities. This is actually the meaning of the summation in the property of the delta in eq.~\ref{delta-prop}:
\ba
    \delta[f(p_N)] = \dfrac{\delta(p_N-p_N^+)}{f_{rec}^+} + \dfrac{\delta(p_N-p_N^-)}{f_{rec}^-}\,.
\ea  
where $f_{rec}^\pm \equiv |\partial f(p_N)/\partial p_N|^{p_N^\pm}$. 

To learn how to proceed in a general case, we will continue our derivations for the two $p_N$ solutions, though, in our case, only one has physical sense and the other solution must be disregarded ($p_N^-$ is negative). 
Putting everything together, and using $d\np_N=p_N^2dp_Nd\Omega_N$, we get:
\ba
    \frac{d^5\sigma(\varepsilon_i)}{d\nk_f d\Omega_N} &=& \frac{{\cal N}}{4/3\pi p_F^3} \int{ p_N^2 dp_N\, \Theta(p_N-p_F) \Theta(p_F-p)}\non\\
    &\times& \dfrac{K}{(2\pi)^2 } L'_{\mu\nu}H^{\mu\nu} \left[\frac{\delta(p_N-p_N^+)}{f_{rec}^+} +\frac{\delta(p_N-p_N^-)}{f_{rec}^-}\right] \,.
\ea
Performing the integral, one obtains:
\ba
    \frac{d^5\sigma(\varepsilon_i)}{d\nk_f d\Omega_N} = \left(\frac{d^5\sigma(\varepsilon_i)}{d\nk_f d\Omega_N}\right)_{p_N^+}
    + \left(\frac{d^5\sigma(\varepsilon_i)}{d\nk_f d\Omega_N}\right)_{p_N^-}\,,
\ea
where each contribution is 
\ba
    \left(\frac{d^5\sigma(\varepsilon_i)}{d\nk_f d\Omega_N}\right)_{p_N^\pm} = 
    \frac{3{\cal N}}{4\pi p_F^3} \dfrac{p_N^2}{f_{rec}} \, \Theta(p_N-p_F) \Theta(p_F-p)\,
    \dfrac{K}{(2\pi)^2 } L'_{\mu\nu}H^{\mu\nu} \,. \label{xs-gRFG}
\ea

Alternatively, we can write:
\ba
    \left(\frac{d^5\sigma(\varepsilon_i)}{d\nk_f d\Omega_N}\right)_{p_N^\pm} = 
    \dfrac{p_N^2}{f_{rec}} \,
    \dfrac{K}{(2\pi)^2 } L'_{\mu\nu}H^{\mu\nu} \,\Theta(p_N-p_F) n_{gRFG}(p). \label{xs-gRFG2}
\ea

\noindent\rule{4cm}{0.4pt}

\subsection{The inclusive cross section}

The inclusive cross section corresponds to the case in which only the final lepton is detected, therefore, one integrates over all (unmeasured) nucleon variables:
\ba
    \frac{d^3\sigma(\varepsilon_i)}{dk_f d\Omega_f} =  k_f^2 \int d\Omega_N{ \left[ \left(\frac{d^5\sigma(\varepsilon_i)}{d\nk_f d\Omega_N}\right)_{p_N^+}
    + \left(\frac{d^5\sigma(\varepsilon_i)}{d\nk_f d\Omega_N}\right)_{p_N^-}  \right]}\,.
\ea
Using the relation $\varepsilon_fd\varepsilon_f = k_fdk_f$ and $-d\omega=d\varepsilon_f$~\footnote{Remember that in the change of variable, we need the absolute value of the Jacobian, i.e., $|{\cal J}| = |\partial\varepsilon_f/\partial\omega|=1$.}, one gets:
\ba
    \frac{d^3\sigma(\varepsilon_i)}{d\omega d\Omega_f} = \dfrac{\varepsilon_f}{k_f}\frac{d^3\sigma(\varepsilon_i)}{dk_f d\Omega_f}\,.
\ea

An interesting property of the RFG model is that it is possible to obtain analytical expressions for the inclusive cross section, i.e., the integrals $\int d\Omega_N$ can be done analytically. This is exercise 16.1 in~\cite{Foundations17}, but we can say that it is highly non-trivial.
If you prefer to invest your time doing something else, you could believe the final results (shown e.g. in \cite{Alberico88}); or do the integrals numerically, actually, current computers can do them in a blink. 
Of course, the same results should be obtained by using the analytical expressions as by doing the integrals numerically. This is a good way of verifying that everything is correct.

\begin{figure}[htbp]
\centering
\includegraphics[width=1.\textwidth,angle=0]{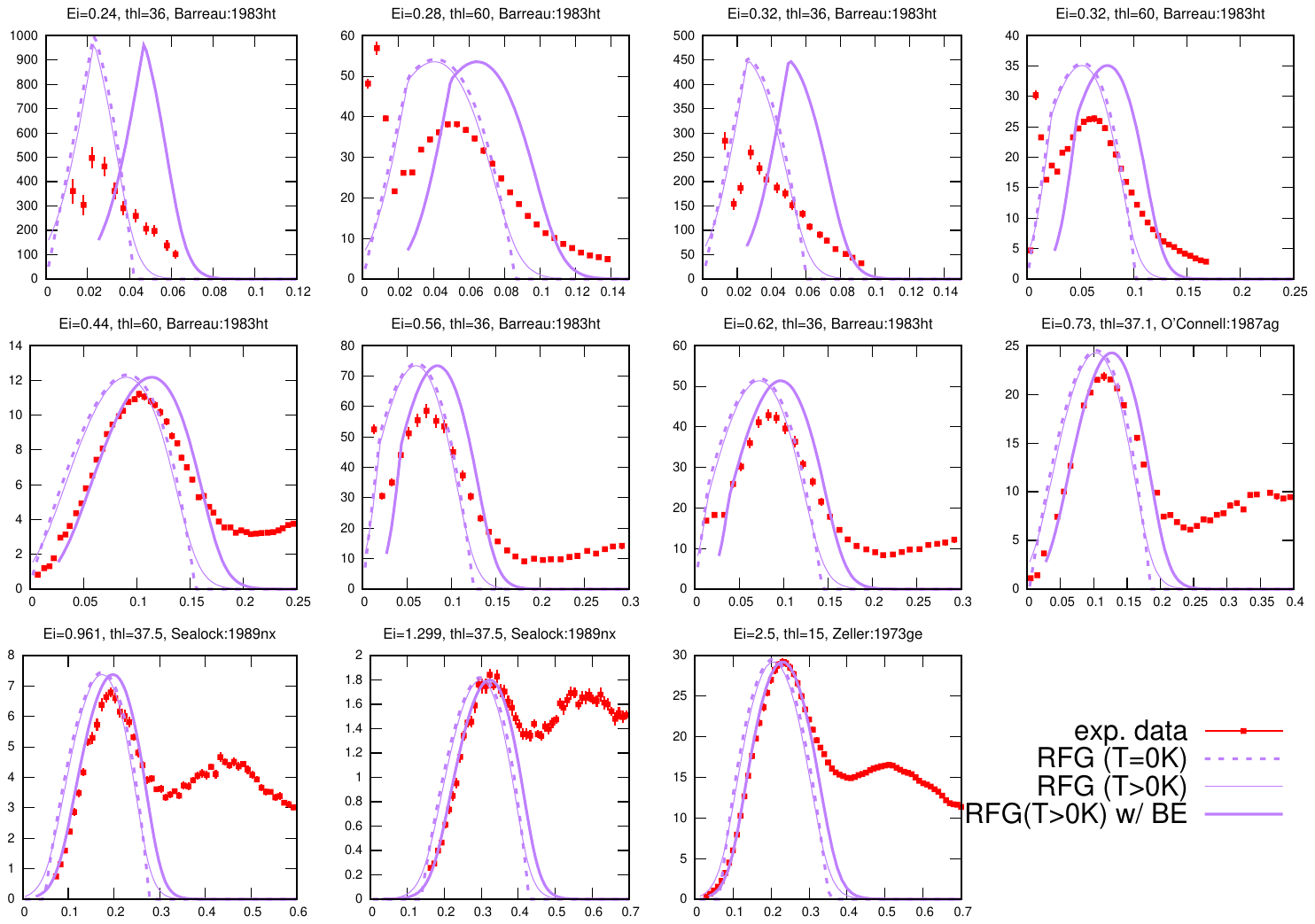}
\vspace{-1.5cm}
\caption{The inclusive cross section [$d\sigma/(d\omega d\Omega_f)$ in units of nb$/$(srad MeV)] for electron-$^{12}$C scattering is presented as a function of the energy transfer $\omega$ (in GeV).
Experimental data is compared with different predictions of the gRFG model for the QE channel: RFG(T=0) is the gRFG with a step-function momentum distribution, RFG(T$>$0) is the gRFG with the Pauli-Dirac momentum distribution, and `w/ BE' means `with binding energy'. Data taken from~\cite{QEarchive}.}\label{fig:RFG}
\end{figure}
In Fig.~\ref{fig:RFG} the inclusive cross section measured for $^{12}$C, for different kinematics, is compared with our theoretical predictions of the QE channel using the RFG model. The experimental data were taken from the data base~\cite{QEarchive}.\\

{\bf When $T>0$ K}, the momentum distribution is more realistic. In the cross section formula, we replace ${\cal N}\frac{\Theta(p_F-p)}{4/3\pi p_F^3}$ by the momentum distribution for $T>0$ K of eq.~\ref{nRFG}. 
For example, for the case shown in Fig.~\ref{fig:mom-dist}, i.e., $p_F=228$ MeV and $k_BT=8$ MeV one obtains
\ba
    {\cal I}(T) = 1.12 \left(\frac{4}{3}\pi p_F^3\right)\,.
\ea

The results for the inclusive cross section with the gRFG at $T>0$ K are in Fig.~\ref{fig:RFG}. Compared to the results at $T=0$ K, one observes that the cross section does not present sharp starting and end points, but it got some small tails in the low- and high-$\omega$ parts of the distribution. As you may have guessed, these tails can be made larger by increasing $T$.

We point out that in the case of $T>0$ K, it is not clear how to consistently implement Pauli blocking. Therefore, for the results in Fig.~\ref{fig:RFG}, we have done the same as for the zero-temperature case, i.e., we have used the sharp step function $\Theta(p_N-p_F)$. \\

Finally, to account for the {\bf binding energy}, we shift the distributions by 24 MeV, which is an average of the binding energies in $^{12}$C. This is an effective way of accounting for the fact that, for knocking out a nucleon, some energy has to be invested in {\it paying} the binding energy. 
The results are shown in Fig.~\ref{fig:RFG}.

\section{Quasielastic scattering within the relativistic local Fermi gas}\label{sec:lRFG}

The so-called local relativistic Fermi gas (lRFG) is another approach widely used in the neutrino-interaction community. It is a variant of the global Fermi gas model described in previous section.
In this case, the nuclear density is taken to be a function of $r$ (the nuclear radius), i.e., $\rho=\rho(r)$ \footnote{Remember that in the global Fermi gas studied in Sec.~\ref{sec:gRFG} the Fermi momentum was a constant free-parameter of the model, consequently, the density was also a constant.}. The Fermi momentum is related to the nuclear density by (see eq.~\ref{rho_pF}):
\ba
  p_F(r) = \left[ 3\pi^2\rho(r) \right]^{1/3}\,,\label{pFr}
\ea
where $\rho$ is normalized to the number of protons or neutrons: $\int \de\nr \rho(r) = {\cal N}$.\\

The local density approximation consists in doing an average of the gRFG cross section using a nuclear density $\rho(r)$ as PDF, i.e. $\int \de\nr\frac{\rho(r)}{{\cal N}}(...)$; and using the Fermi momentum given by eq.~\ref{pFr}. The lRFG cross section reads:
\ba
   \frac{d^5\sigma(\varepsilon_i)}{d\nk_f d\Omega_N}
   &=& \int\de\nr \frac{\rho(r)}{{\cal N}}
    \dfrac{p_N^2}{f_{rec}} \, \Theta\left[p_N-p_F(r)\right] \left[{\cal N}\frac{\Theta\left[p_F(r)-p\right]}{\frac{4}{3}\pi p_F(r)^3}\right] \,
    \dfrac{K}{(2\pi)^2 } L'_{\mu\nu}H^{\mu\nu}\,.
\ea
Considering that the only quantity that depends on $r$ is the Fermi momentum, we get:
\ba
   \frac{d^5\sigma(\varepsilon_i)}{d\nk_f d\Omega_N}
   &=& \dfrac{p_N^2}{f_{rec}} \dfrac{K}{(2\pi)^2 } L'_{\mu\nu}H^{\mu\nu} 
   \int\de\nr \frac{\rho(r)}{{\cal N}}
    \, \Theta\left[p_N-p_F(r)\right] \left[{\cal N}\frac{\Theta\left[p_F(r)-p\right]}{\frac{4}{3}\pi p_F(r)^3}\right]\,.
\ea
In the expression above, we identify the lRFG momentum distribution as:
\ba
 n_{lRFG}(p) = \int\de\nr \frac{\rho(r)}{{\cal N}} \left[{\cal N}\frac{\Theta\left[p_F(r)-p\right]}{\frac{4}{3}\pi p_F(r)^3}\right]\,.\label{np_lRFG1}
\ea
It can be shown (Appendix~\ref{app:norma}) that this momentum distribution is normalized to the number of nucleons, as it should be, i.e., $\int\de\np\, n_{lRFG}(p)={\cal N}$. 
Simplifying eq.~\ref{np_lRFG1} using the definition of $p_F(r)$ in eq.~\ref{pFr}, one gets:
\ba
  n_{lRFG}(p) = \frac{1}{\pi^2}\int\de r r^2 \Theta\left[p_F(r)-p\right]\,.\label{np_lRFG}
\ea
This momentum distribution is plot in Fig.~\ref{fig:mom-dist}, where we have used the nuclear density for protons in $^{16}$O computed with the RMF model of Ref.~\cite{Horowitz81, Sharma93}. The momentum distribution shows a sharp peak at $p=0$, a pronounce decrease for increasing $p$ up to $p\approx30$ MeV, and a soft negative slope for large $p$. The function is zero at $p=p_F(r_{max})$~\footnote{Let's denote {\it the maximum Fermi momentum} by $p_F(r_{max})$, where $r_{max}$ is the point where the density $\rho(r)$, and therefore $p_F(r)$, are maxima.}. 

Reordering the terms and performing the trivial integral over the solid angle, the lRFG cross section reads:
\ba
   \frac{d^5\sigma(\varepsilon_i)}{d\nk_f d\Omega_N}
   &=& \dfrac{p_N^2}{f_{rec}} \dfrac{K}{(2\pi)^2 } L'_{\mu\nu}H^{\mu\nu} 
   \frac{1}{\pi^2}\int\de r r^2 
    \, \Theta\left[p_N-p_F(r)\right]\Theta\left[p_F(r)-p\right]\,
\,.
\ea
$\Theta\left[p_N-p_F(r)\right]$ is the usual Pauli blocking (PB) factor, which will give something similar to what is discussed in Section~\ref{PB-lFG}. 

\begin{figure}[htbp]
\centering  
\includegraphics[width=.5\textwidth,angle=0]{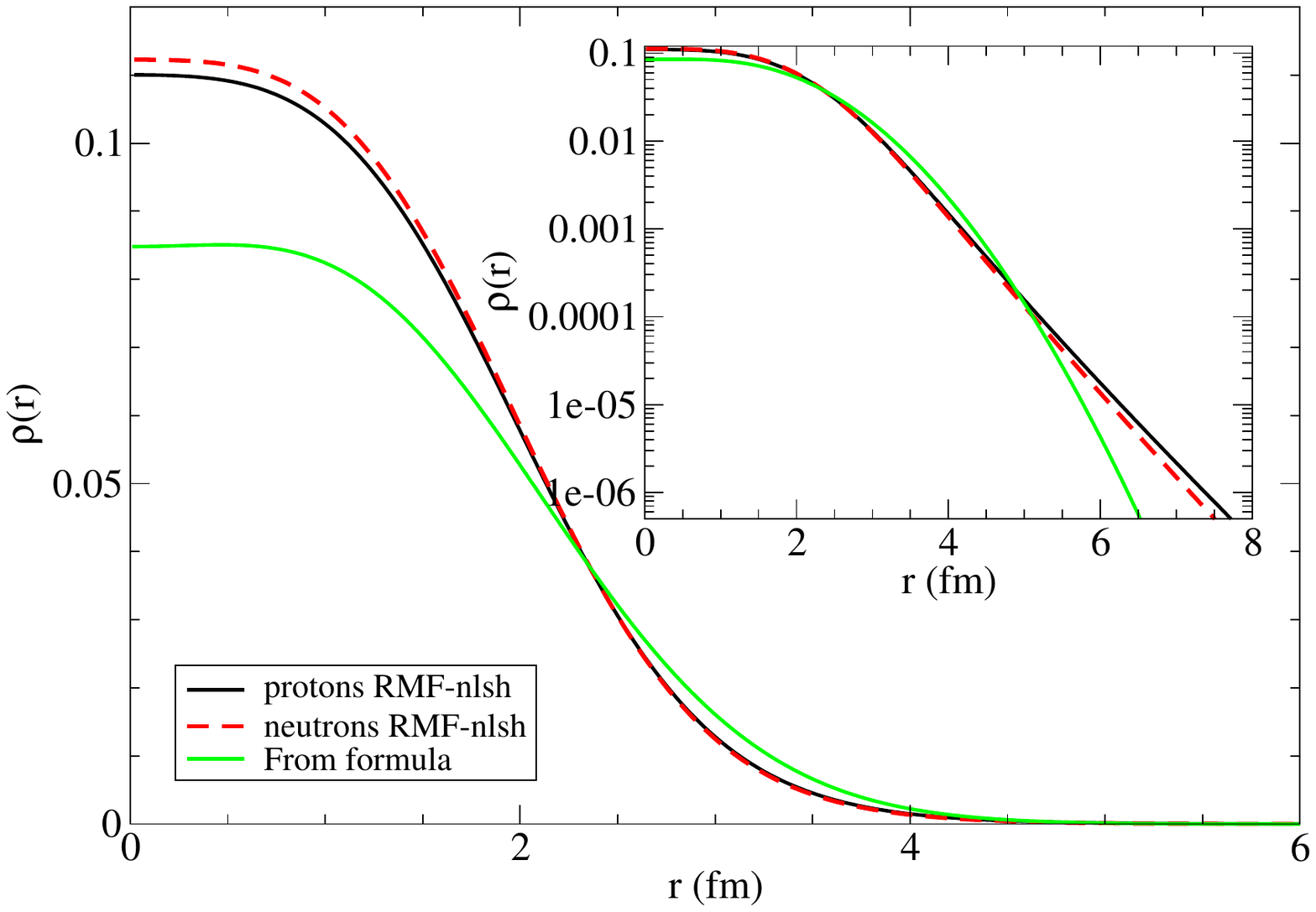}
\includegraphics[width=.5\textwidth,angle=0]{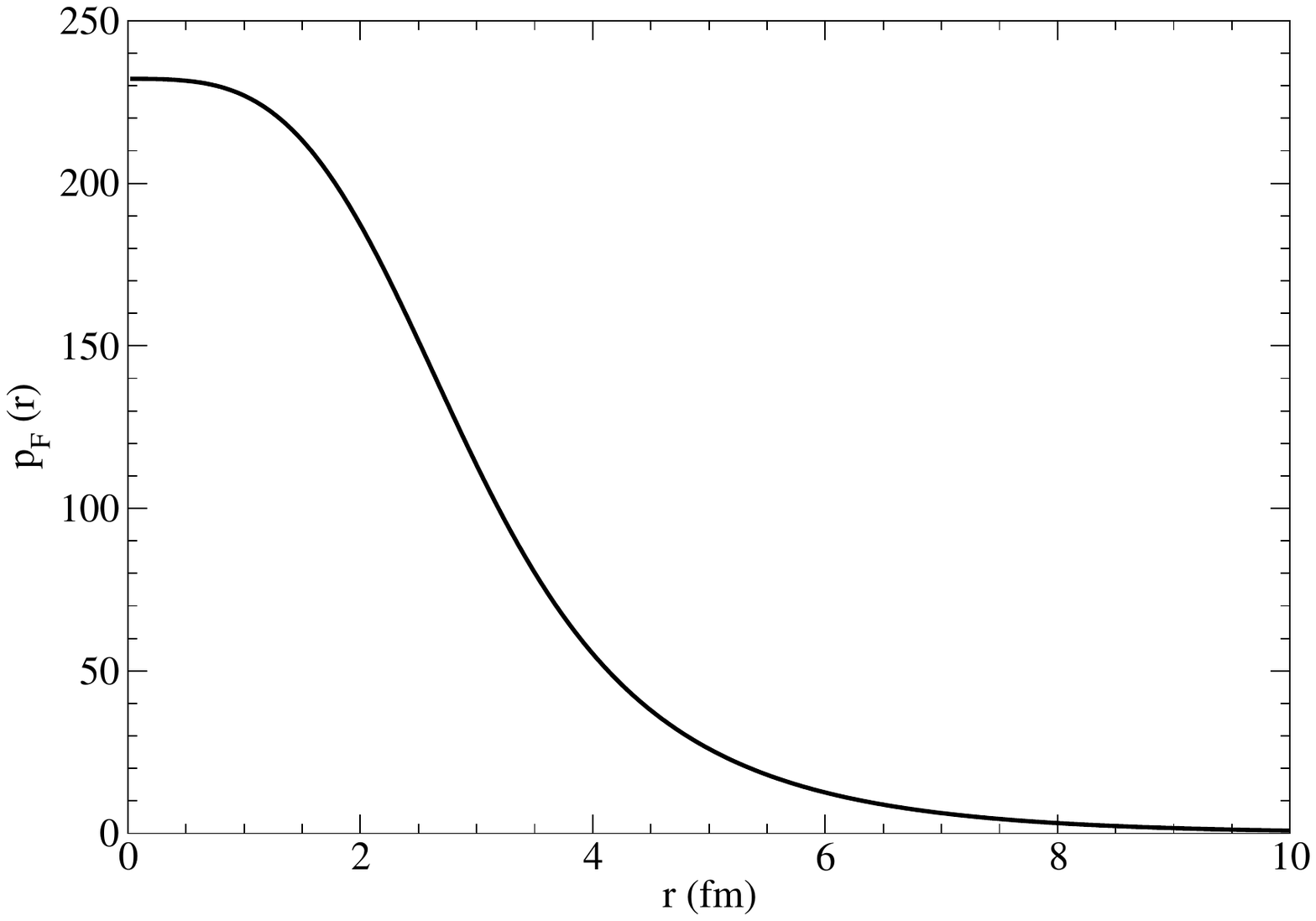}
\vspace{-0.5cm}
\caption{Left panel: $^{12}$C proton and neutron densities computed with the RMF-NLSH model~\cite{Horowitz81, Sharma93}. Solid green is the density using eq.~\ref{rho_analy} and the parameters of eq.~\ref{c12_param}. $\rho$ is in fm$^{-3}$. The smaller plot inside is the same but in log-scale. Right panel: Fermi momentum as function of $r$, computed as eq.~\ref{pFr}. $p_F$ is in MeV. The proton density of the RMF-NLSH model was used.}
\label{fig:rho_and_pF}
\end{figure}

Let us study the result of the integral over $r$ in the equation above.  
For that, let's take a look to the different pieces that enter this equation. 
In the left panel of Fig.~\ref{fig:rho_and_pF} we show the nuclear density of protons and neutrons $\rho(r)$ as a function of $r$. In the right panel of Fig.~\ref{fig:rho_and_pF} we show the Fermi momentum, derived from these densities using eq.~\ref{pFr}. For the nuclear density we have used the RMF-NLSH~\cite{Sharma93} model, and a simpler analytical expression obtained by parametrizing experimental data for the proton densities and theoretical results for the neutron densities~\cite{LeitnerThesis}. The expression for the analytical densities and related useful results are in appendix~\ref{app:NuclearDensity}.

For simplicity, let's assume that we are above the PB region, so $p_N$ is larger than the maximum Fermi momentum, $p_F(r_{max})$. 
Thus, the cross section reads:
\ba
   \frac{d^5\sigma(\varepsilon_i)}{d\nk_f d\Omega_N}
   &=& \dfrac{p_N^2}{f_{rec}} \dfrac{K}{(2\pi)^2 } L'_{\mu\nu}H^{\mu\nu} 
   n_{lRFG}(p)\,.\label{xs_lRFG_noPB}
\ea

\begin{figure}[htbp]
\centering
\includegraphics[width=1.\textwidth,angle=0]{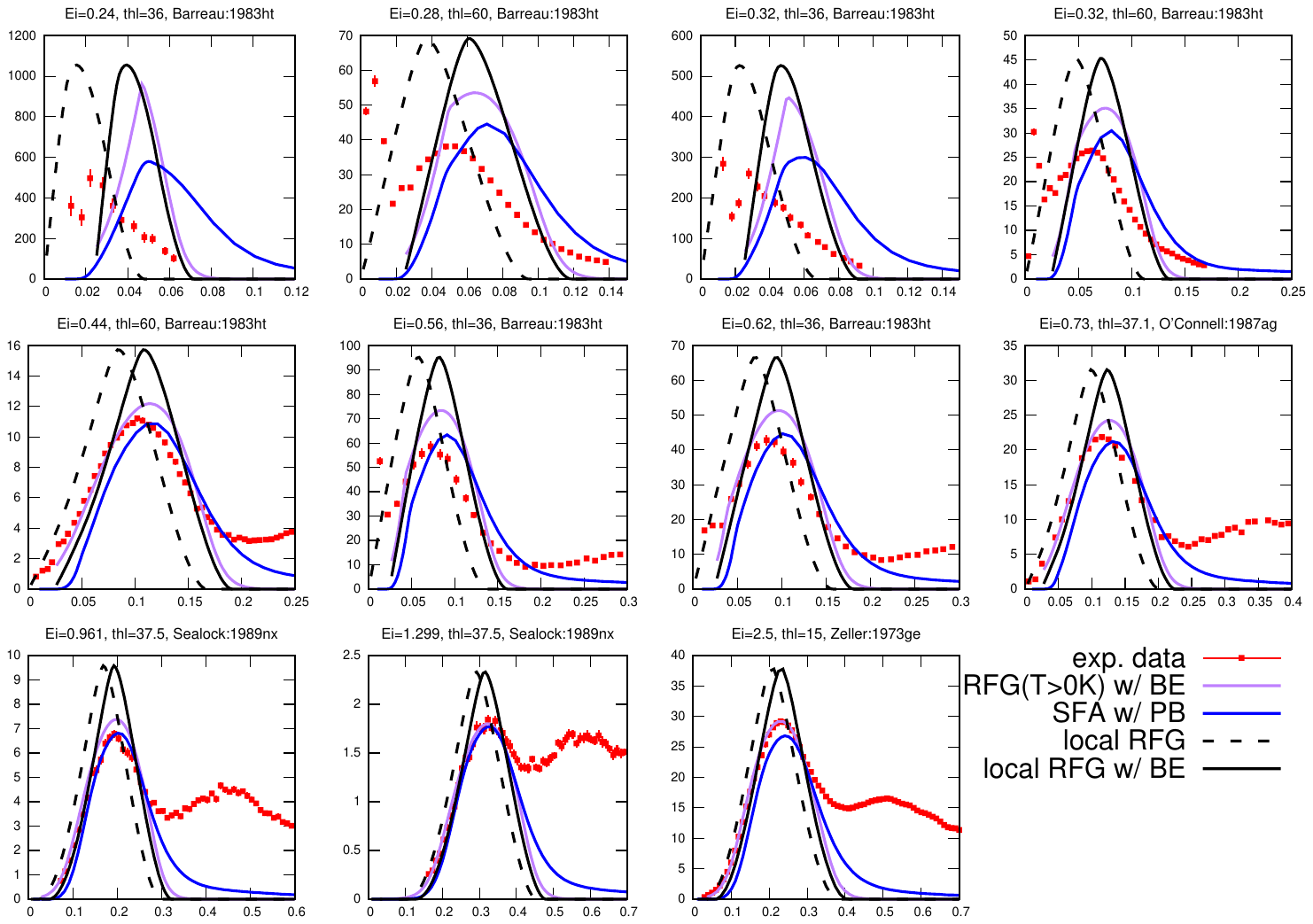}
\vspace{-1.5cm}
\caption{As figure~\ref{fig:RFG} but now for the models: RFG(T$>$0) with binding energy (BE), the SFA with Pauli Blocking (Section~\ref{sec:SFA}), the lRFG and the lRFG with BE. 
}\label{fig:lRFG}
\end{figure}

The cross sections are shown in Fig.~\ref{fig:lRFG}. With respect to the inclusive cross section computed with the gRFG model, the strength in the lRFG is redistributed, increasing it when $p=0$ MeV (namely, at the center of the QE peak) and adding tails to the distributions for large $p$ (namely, at small and large $\omega$ values). 
Comparing eqs.~\ref{xs-gRFG2} and \ref{xs_lRFG_noPB}, it is easy to understand that these differences in the cross section results are simply explained by the differences in the momentum distributions of the gRFG and lRFG.

\section{Quasielastic scattering within the PWIA with a realistic spectral function}\label{sec:SFA}

We study now the cross section formulas for the case of `Plane-Wave Impulse Approximation (PWIA) with a realistic spectral function', we will also refer to this as `factorized Spectral Function Approach' (SFA). 
For further details on this approach, see the review article~\cite{Benhar08}; also recommended is the reading of Ref.~\cite{VanOrden19}, where the one-hole spectral function is obtained within the gRFG, lRFG and the PWIA approaches in the context of CCQE neutrino-nucleus scattering.  

We start with eq.~\ref{d6sig}, i.e.:
\ba
    d^6\sigma(\varepsilon_i) = K  (2\pi)^4\delta^4(K_f+P_N-K_i-P) L'_{\mu\nu}H^{\mu\nu}\dfrac{d\nk_f}{(2\pi)^3} \dfrac{d\np_N}{(2\pi)^3}\,.\non
\ea
As in the gRFG case, we average over initial nucleon 4-momentum $P$, i.e.: 
\ba
  d^6\sigma(\varepsilon_i) &=& \int d\np \int dE\ S(E,\np)\non\\
  &\times& K  (2\pi)^4\delta^4(K_f+P_N-K_i-P) L'_{\mu\nu}H^{\mu\nu}\dfrac{d\nk_f}{(2\pi)^3} \dfrac{d\np_N}{(2\pi)^3}\,,\label{SFd6sig}
\ea
where the spectral function $S(E,\np)$ is the analogous quantity to the factor $\delta(E-\sqrt{p^2+M^2})\frac{{\cal N}}{4/3\pi p_F^3}\Theta(p_F-p)$ in the gRFG, i.e., it is the PDF of finding a nucleon with energy $E$ and momentum $\np$ in the nucleus.
It is defined in the lab frame (target nucleus at rest), therefore, since the Fermi motion is isotropic, one has $S(E,\np)\rightarrow S(E,p)$. 
Note that within this model, and contrary to the RFG cases, the bound nucleons are considered off-shell, i.e., $E^2\neq p^2+M^2$. 

\begin{figure}[htbp]
\centering
\includegraphics[width=.6\textwidth,angle=0]{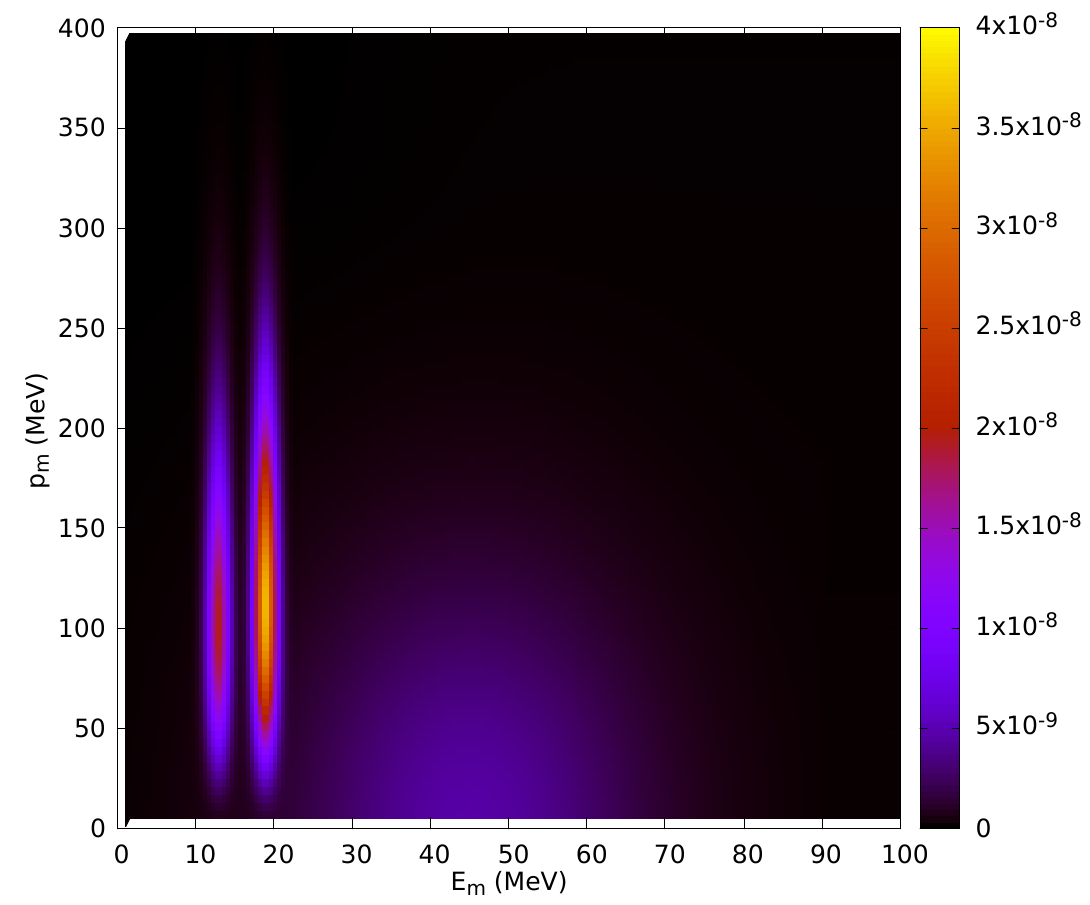}.
\caption{The Rome spectral function for protons in $^{16}$O~\cite{Benhar94,Benhar05} as function of missing energy and missing momentum. 
We can clearly identify the two narrow and low-$E_m$ states $1p_{3/2}$ and $1p_{1/2}$, centered at approximately 13 and 19 MeV, respectively. 
The broad $1s_{1/2}$ state, centered at approximately 45 MeV, is also visible. 
The SRC background, not visible in this plot, could be highlighted using log-scale in the z axis (see e.g.~\cite{VanOrden19}).}\label{fig:Rome-SF}
\end{figure}

The spectral function is normalized to the number of active nucleons ${\cal N}$, which can be the number of protons, neutrons, or both:
\ba
 \int d\np \int dE\ S(E,p) = {\cal N}\,.
\ea 
 The momentum distribution is given by:
\ba
  n_{SFA}(p) = \int dE S(E,p)\,. 
\ea
As an example, in Fig.~\ref{fig:Rome-SF} we show the spectral function for $^{16}$O from Refs.~\cite{Benhar94,Benhar05}. 
It is constructed as the sum of two terms:  
\ba 
S(E_m,p) = S_{\text{mean-field}}(E_m,p) + S_{SRC}(E_m,p)\,,
\ea
where
\ba
 S_{\text{mean-field}}(E_m,p) = \sum_\kappa n_\kappa(p)\rho_\kappa(E_m),
\ea
with $n_\kappa(p)$ the momentum distribution for an energy level $\kappa$ and $\rho_\kappa(E_m)$ is usually a Gaussian (or similar) distribution normalized to 1 for the case of fully occupied states. The normalization of this $\rho_\kappa(E_m)$ function is called {\bf spectroscopic factor}.

Experimentally, one observes a depletion of the occupation of the mean-field levels with respect to their value in the independent-particle shell model, attributed to effects beyond mean field (long- and short-range correlations). An effective and simple way to somehow account for this is to use constant spectroscopic factors (of course, always lower than 1).

The factor $S_{SRC}$ is a contribution that accounts for the number of nucleons that are in the nucleus as short-range correlated pairs (approximately 20\%).

There are different approaches to construct the function $S(E,p)$: from microscopic ab initio models, to other semi-phenomenological approaches that combine experimental information (mainly from electron scattering experiments) with mean-field based models. 
We do not discuss any of these approaches in these notes. 
For the purpose of this course,
it is enough to consider that $S(E,p)$
is a realistic representation of the probability of removing a particle with momentum $p$ and leaving the resulting $(A-1)$ system with an excitation energy $E$. 
For further reading see e.g.~\cite{Alvarez-Ruso25} and references therein.

The {\bf missing energy} $E_m$ is defined as the energy that the residual system receives and is not kinetic energy, i.e.: 
\ba
  E_m \equiv \omega - T_N - T_B = M_B + M - M_A\,.
\ea
$T_N$ and $T_B$ are the kinetic energies of the final nucleon and residual system, $M_A$ and $M_B$ are the masses of the target nucleus and the residual system~\footnote{Note that the residual system could be the nucleus with $A-1$ nucleons in the ground or an excited state, which is the easier scenario, but it could also be an unbound system of $A-1$ nucleons and other particles. So, in general, within this model we do not specify any information about residual system, but only its 4-momentum $P_B$ which is determined by energy-momentum conservation.}.
Comparing this equation with the energy conservation equation in the nucleonic vertex, i.e.:
\ba
  \omega + E = E_N\,,
\ea
one gets the following relation between $E_m$ and $E$:
\ba
  E = M - E_m - T_B\,,\label{defE}
\ea
which can be used as definition of the energy of the initial nucleon (which is needed for the calculation of the matrix element).
In the literature, one finds  alternative definitions of the energy of the bound nucleon, $E$; we do not discuss it here, but point out that, within this approach, bound nucleons are off-shell, i.e.:  $E^2 \neq p^2+M^2$. The different prescriptions or models used to defined $E$ are sometimes referred as off-shell ambiguities.
This off-shell character of the bound nucleon implies that there is an additional degree of freedom (or one less constraint) compared to the Fermi gas approaches. 

In what follows, we will neglect the nuclear recoil $T_B$, in other words, we neglect the kinetic energy of the residual system with respect to its mass ($E_B\approx M_B$). 
This is a good approximation that simplifies all the kinematic relations. We use the 4-momentum conservation Dirac delta in eq.~\ref{SFd6sig} to integrate over $E$ and $\np$; thus, one gets:
\ba
  \dfrac{d^6\sigma(\varepsilon_i)}{d\nk_f d\np_N} = S(E_m,p) \, \frac{K}{(2\pi)^2} L'_{\mu\nu}H^{\mu\nu}\,,\label{SFd6sig2}
\ea
where in the SF we have replaced $E$ by $E_m$.
Hence, the six-differential cross section in terms of laboratory variables reads:
\ba
  \dfrac{d^6\sigma(\varepsilon_i)}{d\Omega_f dk_f d\Omega_N dp_N} = S(E_m,p) \, p_N^2k_f^2\frac{K}{(2\pi)^2} L'_{\mu\nu}H^{\mu\nu}\,.\label{SFd6sig3}
\ea
The lepton and hadron tensors are the single-nucleon ones derived in Section~\ref{sec:elastic}~\footnote{Watch out! Attention must be paid of making the derivation for off-shell nucleons, i.e., making sure that on-shell relations are not used on the way to get the expressions of the hadron tensor.}.

Sometimes, it is is useful to express the differential cross section in terms of other kinematic variables. For example, the differential cross section as function of the missing energy gives us direct information on where the knocked out nucleon came from. 
To change from one independent variable to another, we use the Jacobian, in this case:  ${\cal J}=\left|\frac{dE_m}{dp_N}\right|$, i.e., we need the relation between $E_m$ and $p_N$. From energy-momentum conservation and neglecting the nuclear recoil we have:
\ba
 E_m \approx \omega - T_N \Longrightarrow dE_m = -dE_N\,.
\ea
Hence,
\ba
  {\cal J} = \left|\frac{dE_m}{dp_N}\right| = \left|\frac{dE_m}{dE_N}\frac{dE_N}{dp_N}\right| = \frac{p_N}{E_N}\,. \label{Jacobian_easy}
\ea
If we do not neglect the nuclear recoil, the Jacobian is a bit more complicated (see Appendix~\ref{app:Jacobian}).

Finally, the cross section as a function of the missing energy is:
\ba
  \dfrac{d^6\sigma(\varepsilon_i)}{d\Omega_f dk_f d\Omega_N dE_m} = S(E_m,p) \, {\cal J}^{-1} p_N^2k_f^2\frac{K}{(2\pi)^2} L'_{\mu\nu}H^{\mu\nu}\,.\label{SFd6sig4}
\ea

We point out that if one uses $E_m$, instead of $p_N$, as independent variable, $p_N$ must be obtained from energy-momentum conservation. If nuclear recoil is not neglected, one finds a quadratic equation, with two possible solutions for $p_N$. 
In this case, there is a small region of phase space where one should consider the two possible solutions; this is discussed in detail in Appendix B of Ref.~\cite{Nikolakopoulos25}.

\begin{figure}[htbp]
\centering  
\includegraphics[width=1.\textwidth,angle=0]{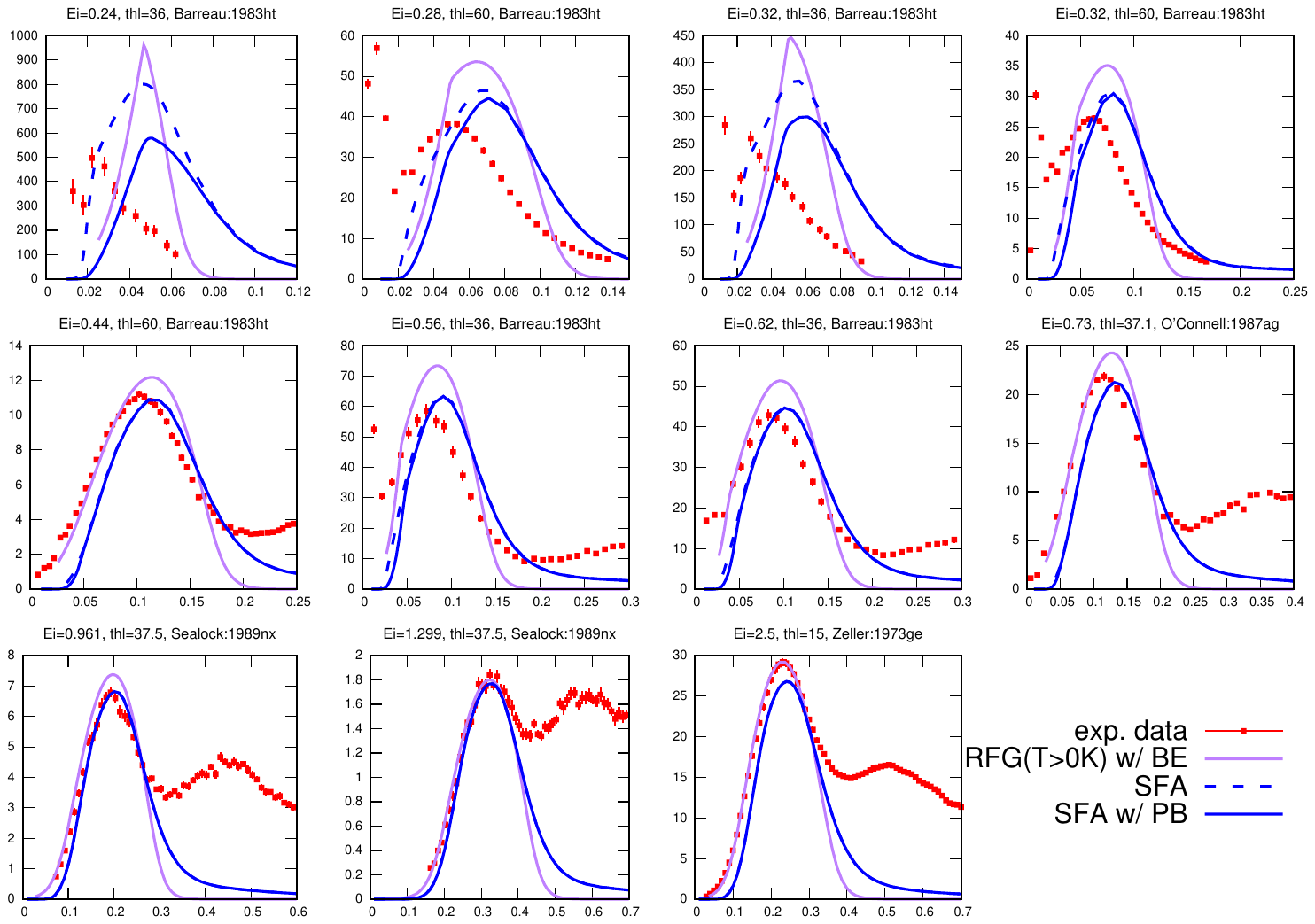}
\vspace{-1.5cm}
\caption{As figure~\ref{fig:RFG} but now for the models: RFG(T$>$0) with binding energy, the SFA and the SFA with Pauli Blocking (PB). 
}\label{fig:SFA}
\end{figure}

\subsection{Pauli blocking}\label{PB-lFG}

We can implement Pauli's exclusion principle in a relatively simple way, similar to what is done in the local Fermi gas model (Section~\ref{sec:lRFG}). 
The method consists in multiplying the cross section by a {\it Pauli-blocking factor}, which is a function of the momentum of the final nucleon:
\ba
  \left.\dfrac{d^6\sigma(\varepsilon_i)}{d\nk_f d\np_N}\right|_{\text{Pauli blocked}} = \dfrac{d^6\sigma(\varepsilon_i)}{d\nk_f d\np_N} \left(1 - \frac{1}{{\cal N}}\int{d\nr\ \rho(r)\ \theta[p_F(r) - p_N] } \right)\,,
\ea
where
$p_F(r) = \left[ 3\pi^2 \rho(r) \right]^{1/3}$,
${\cal N}$ is the number of protons or neutrons and $\rho$ is the proton or neutron density normalized to the number of nucleons, i.e. $\int d\nr \rho(r) ={\cal N}$. 

\begin{figure}[htbp]
\centering  
\includegraphics[width=.5\textwidth,angle=0]{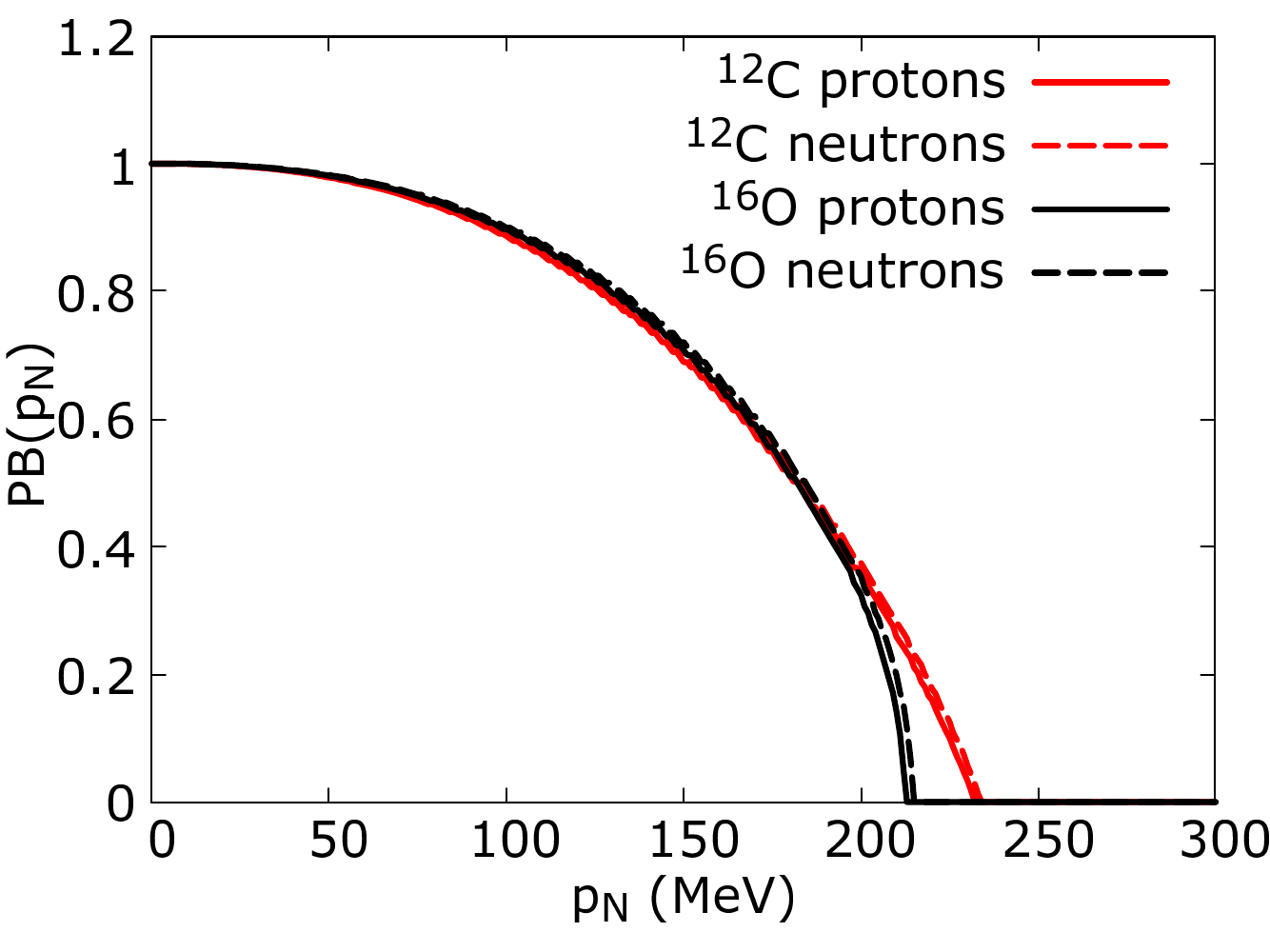}
\caption{Pauli blocking function within the lRFG computed with the barionic densities from an RMF model for oxygen and carbon (see text).}\label{fig:PB_LFG}
\end{figure}

We define a Pauli blocking function that depends on $p_N$:
\ba
  PB(p_N) \equiv  \frac{1}{{\cal N}} \int{d\nr\ \rho(r)\ \theta[p_F(r) - p_N] }\,.
\ea
It can be computed numerically, for different nuclei, and stored in a file. It is represented in Fig.~\ref{fig:PB_LFG} for carbon and oxygen, the barionic density from a relativistic mean-field (RMF) model~\cite{Walecka74,Horowitz81,TIMORA,Sharma93} was used. 
When $PB(p_N)$ is maximum (namely, equals $1$), i.e. when $p_N$ is zero, the process is completely blocked, so the cross section is zero; when $PB(p_N)$ is half of its maximum, it reduces the cross section by 50\%, and so on.   

\section{Quasielastic scattering with a RDWIA}

The bound and \underline{scattered} nucleon states are solutions of Dirac equation in  the presence of relativistic mean-field potentials. We refer to such a model as a relativistic distorted-wave impulse approximation (RDWIA). 
See references~\cite{Gonzalez-Jimenez19, VanOrden19, Franco-Munoz25} and references therein for more information.

\begin{figure}[htbp]
\centering  
\vspace{-1.1cm}
\includegraphics[width=1.05\textwidth,angle=0]{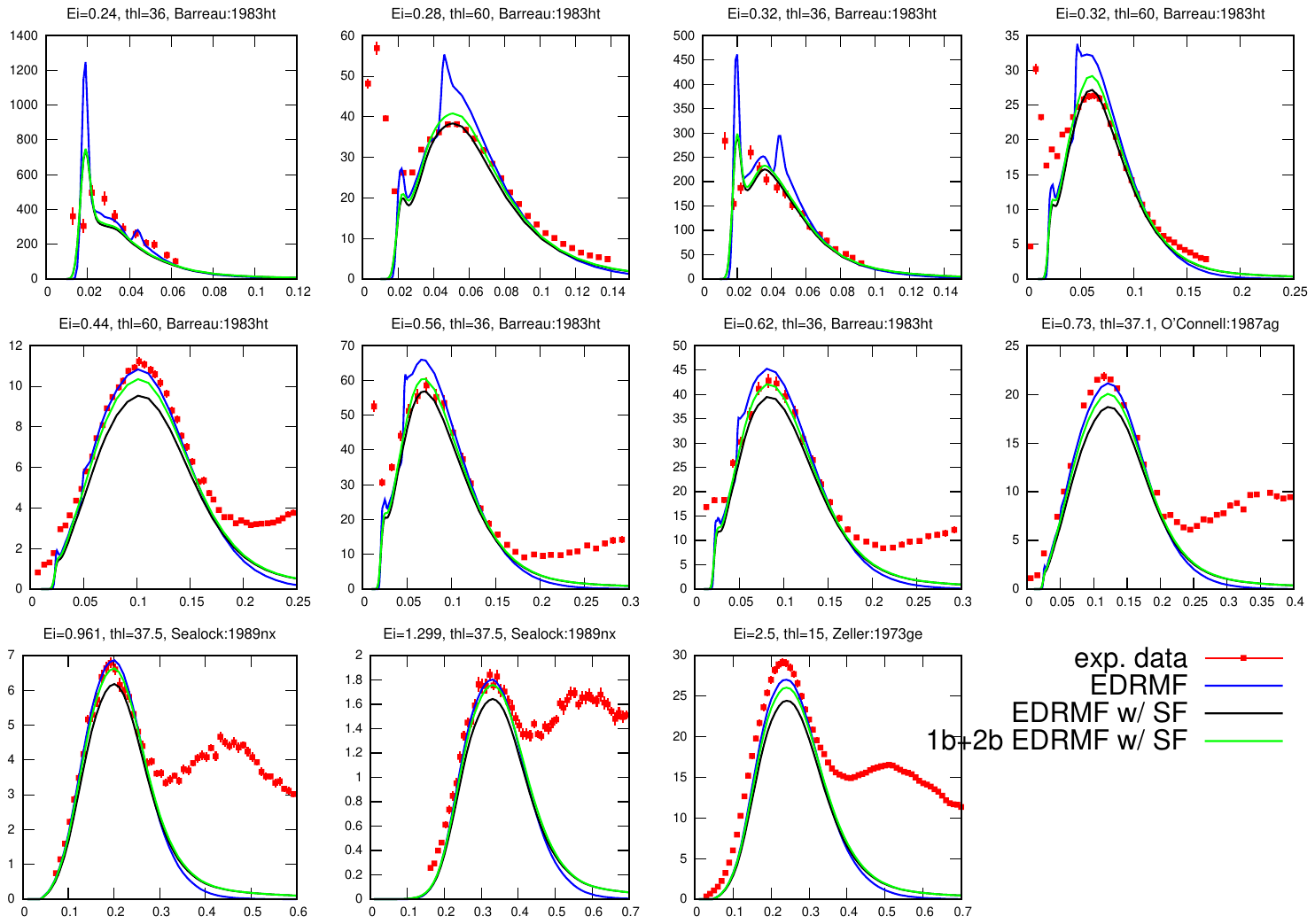}
\vspace{-1.5cm}
\caption{As figure~\ref{fig:RFG} but now for the models: EDRMF, EDRMF with a spectral function (SF), and EDRMF with a SF and using a 1-body and a 2-body current operator. 
}\label{fig:edrmfs}
\includegraphics[width=1.05\textwidth,angle=0]{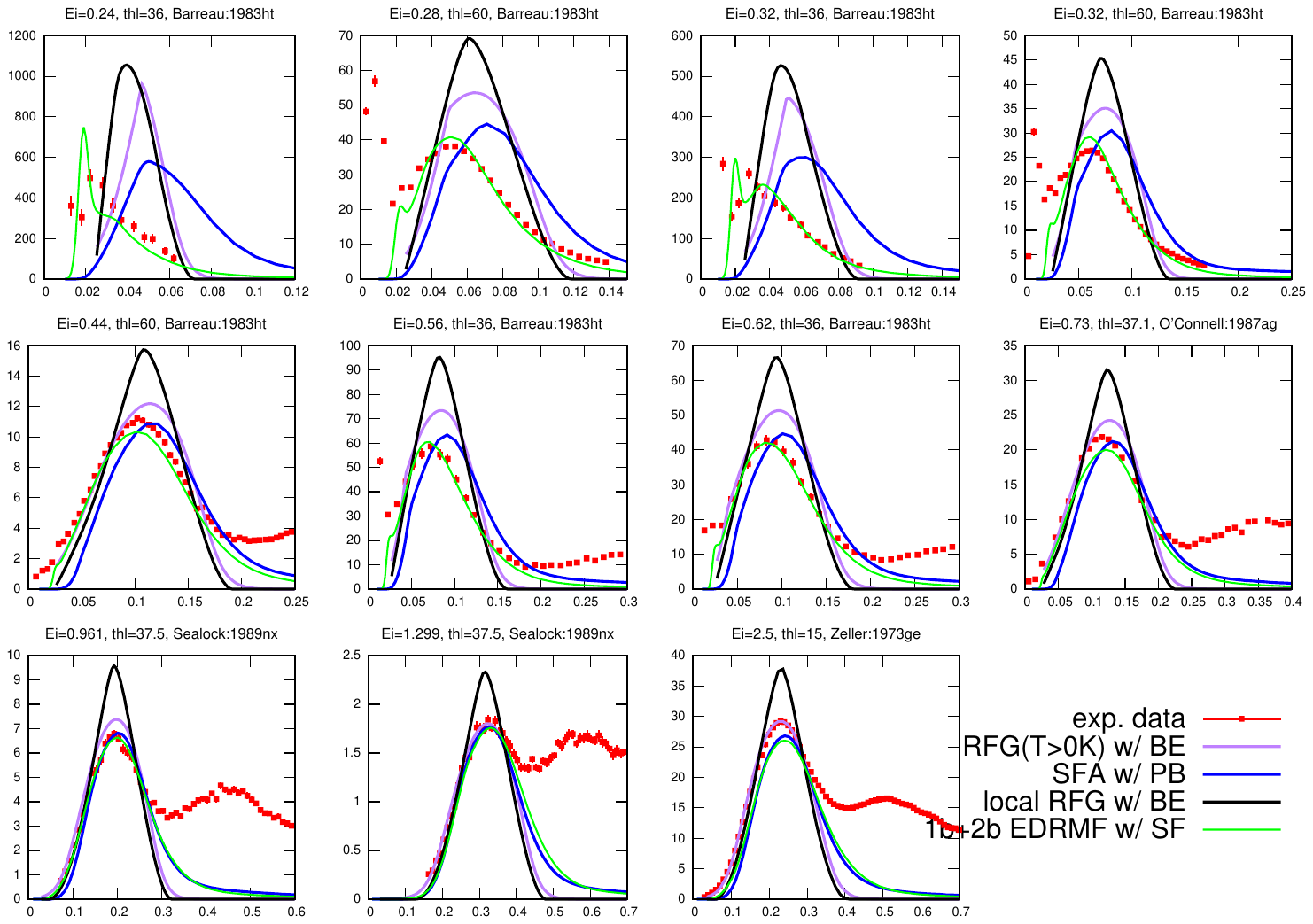}
\vspace{-1.5cm}
\caption{As figure~\ref{fig:RFG}. Model predictions: RFG(T$>$0) with BE, the SFA with Pauli Blocking, the lRFG with BE, and the EDRMF with a SF and 1b+2b current operator. 
}\label{fig:all-final}
\end{figure}

In Fig.~\ref{fig:edrmfs}, we show results with different approaches:
\begin{itemize}
 \item {\bf An independent-particle shell model} (IPSM). The bound and scattered nucleon states are solutions of the Dirac equation in presence of the relativistic mean-field potentials obtained in the Dirac-Hartree approach with the set of parameters from Ref.~\cite{Sharma93}. 
 
 The strong energy independent potentials from the IPSM work well for scattered nucleons with relatively low momentum. The potentials were fitted to reproduce nuclear matter properties, so it is not a surprise that for fast nucleons (above the Fermi level), these potentials do not work well anymore. They are too strong, which causes the cross section to be shifted (too much) to the high $\omega$ part, worsening the agreement with data. The energy-dependent relativistic mean-field approach (EDRMF) was proposed in~\cite{Gonzalez-Jimenez19} to amend that. The potentials are the RMF ones for low nucleon (kinetic) energy but are made softer for increasing energy.  In Fig.~\ref{fig:edrmfs} it is labeled as `EDRMF'. 
\item Incorporating {\bf effects beyond mean field} using a phenomenological spectral function. Long- and short-range nucleon-nucleon correlations are not fully taken into account by the mean field. We replace the fully occupied shells by partially occupied ones. As in the SF approach, the nucleons that disappear from the shells are placed in a background that in a phenomenological way accounts for SRC. In Fig.~\ref{fig:edrmfs}, it is labeled as `EDRMF w/ SF'.
\item Incorporating {\bf effects beyond the impulse approximation}. We do not describe any of these here, just mention that effects beyond the one-body current operator, which is implicit in the impulse approximation, exist and have been studied in the literature. As an example, we show the effect in the QE 1p-1h cross section of including a two-body current operator (original calculations and more details in~\cite{Franco-Munoz23,Franco-Munoz25}). In Fig.~\ref{fig:edrmfs}, it is labeled as `1b+2b EDRMF w/ SF'. 
\end{itemize}
In Fig.~\ref{fig:all-final} we show a compilation of all models discussed in these notes. 

\newpage 

\section{Charged-current quasielastic neutrino-nucleus scattering}

The cross section formula for charged-current (CC) elastic neutrino-nucleon scattering~\footnote{It may be a bit confusing to call 
{\it elastic} the CC reaction 
\ba
\nu_l+n\longrightarrow l + p\,,\label{nuN->lN}
\ea 
the particles in the initial and final state are not the same. However, we use this term for analogy with the electron-nucleon scattering case. 
In the literature, sometimes one finds the term {\it quasielastic} to refer to the reaction in~\ref{nuN->lN}; we intentionally avoid to do so because {\it quasielastic} was already taken to refer to the lepton-nucleus reaction.} and CC quasielastic (CCQE) neutrino-nucleus scattering is obtained following the same steps as for electron scattering. 

In this case, the lepton tensor reads:
\ba
    L'_{\mu\nu} &=& 2\left[ K_{i,\mu} K_{f,\nu} + K_{i,\nu} K_{f,\mu} - g_{\mu\nu}K_i\cdot K_f \pm ih\epsilon_{\mu\nu\alpha\beta}K_i^\alpha K_f^\beta\, \right]\,,
\ea
where the helicity $h=+1$ for antineutrino and $h=-1$ for neutrinos.

The hadron tensor is more complicated due to the vector (V) and axial-vector (A) structure of the currents. The hadronic current operator can be written as:
\ba
    \Gamma^\mu = \Gamma_{V}^\mu + \Gamma_{A}^\mu
\ea
with $\Gamma_{V}^\mu$ and $\Gamma_{A}^\mu$ current operators that account for the internal structure of the nucleon. They are given in Appendix~\ref{app:structure}. 

After some lengthy calculation, the hadron tensor can be expressed as (see~\cite{MegiasPhD} for details):
\ba
    H^{\mu\nu} &=& -W_1g^{\mu\nu}+W_2\frac{P^\mu P^\nu}{M^2} + iW_3\epsilon^{\mu\nu\alpha\beta}\frac{P_\alpha Q_\beta}{2M^2}
    + W_4\frac{Q^\mu Q^\nu}{M^2} + W_5\frac{P^\mu Q^\nu + Q^\mu P^\nu}{2M^2}\,,
\ea
where the structure functions are given by
\ba
    W_1 &=& \tau[(F_1^V+F_2^V)^2 + G_A^2] + G_A^2\,,\\
    W_2 &=& (F_1^V)^2 + \tau(F_2^V)^2 + G_A^2\,,\\
    W_3 &=& 2G_A(F_1^V+F_2^V)\,,\\
    W_4 &=& \frac{(F_2^V)^2}{4} \left(\tau-1\right) - \frac{F_1^VF_2^V}{2} - M F_P G_A + \tau M^2F_P^2\,,\\
    W_5 &=& W_2\,,
\ea
with $\tau\equiv\frac{|Q^2|}{4M^2}$. 
$F_{1,2}^V$ are the vector-isovector form factors of the nucleon while $G_A$ and $F_P$ are the axial-vector (or pseudovector) and axial (or pseudoscalar) nucleon form factors. They are all functions of the squared 4-momentum transfer $Q^2$.

Finally, one has to replace the EM coupling by the weak CC one, which in the limit $\frac{|Q^2|}{M_W^2}<<1$ reads: 
\ba
    \left(\frac{4\pi\alpha}{Q^2}\right)^2\longrightarrow \left(\frac{G_F\cos\theta_c}{\sqrt2}\right)^2\,,
\ea
with $\cos\theta_c$ the Cabibbo angle and $G_F$ the Fermi constant.

\subsection{Neutrino experiments: average over the neutrino energy}

In neutrino experiments, the energy of the incoming neutrino (flux) is only known as a broad distribution. Therefore, one averages over this energy for the given distribution $\phi(\varepsilon_i)$, that depends on the experiment. 

For example, the {\bf elastic lepton-nucleon cross section} as a function of the lepton variables, reads:
\ba
  \langle\frac{d^3\sigma}{d\nk_f}\rangle = \int_{\varepsilon_{min}}^{\varepsilon_{max}} \phi(\varepsilon_i) d\varepsilon_i \frac{K}{(2\pi)^2} \delta(\varepsilon_f + E_N-\varepsilon_i-E) L'_{\mu\nu}H^{\mu\nu}\,,\label{d3sigl_flux}
\ea
with the flux normalized so that $\int{\phi(\varepsilon_i)d\varepsilon_i} = 1$. The symbols $\langle...\rangle$ denote that it is a flux-averaged (or flux-folded) cross section. 
In this particular case, the integral over the neutrino energy can be done analytically using the energy-conservation Dirac delta, so the triple differential cross section reads: 
\ba
  \langle\frac{d^3\sigma}{d\cos\theta_f d\phi_f dk_f}\rangle = \frac{k_f^2}{f_{rec}} \phi(\varepsilon_i) \frac{K}{(2\pi)^2} L'_{\mu\nu}H^{\mu\nu}\,,
\ea
with 
$$f_{rec}=\left| -1 + \frac{k_i-k_f\cos\theta_f}{E_N} \frac{\varepsilon_i}{k_i} \right|\,,$$
and $\varepsilon_i$ given as a function of the independent variables, analogously as done in Appendix~\ref{full-kinematics}.\\

On the other hand, the {\bf quasielastic lepton-nucleus cross section} reads:
\ba
    \langle\frac{d^6\sigma}{d\nk_f d\np_N}\rangle &=& \int_{\varepsilon_{min}}^{\varepsilon_{max}} \phi(\varepsilon_i)d\varepsilon_i\,\, 
    \frac{{\cal N}}{4/3\pi p_F^3} \int_0^{p_F} d^3\np\,\, \Theta(p_N-p_F)\non\\
    &\times& \dfrac{K}{(2\pi)^2}  \delta^4(K_f+P_N-K_i-P) L'_{\mu\nu}H^{\mu\nu}\,.
\ea
As always, one chooses over which variables to analytically integrate using the Dirac delta.

\newpage

\appendix

\section{Conventions}\label{app:Conventions}
  
\textbf{Natural units:} 
\ba
\hbar=c=1\,.\label{hbc}
\ea 
The transformation between natural units and physical units is given by the factor: 
\ba
\hbar c = 197.33\,\text{MeV fm}\,.
\ea

\textbf{4-vectors and 3-vectors:} 4-vectors are represented with capital letters ($A$), and sometimes, for clarity, with a Greek super- or sub-index $A^{\mu}$. 3-vectors are represented in bold, e.g. ${\bf a}$, and their magnitude as $a$.\\

\textbf{The metric} is given by
\begin{equation}
g_{\mu\nu}=\left(\begin{array}{rrrr}
1 &  &  &  \\
 & -1 &  &  \\
 &  & -1 &  \\
 &  &  & -1
\end{array}\right)=
\left(\begin{array}{cc}
1 &  \\
 & -\munit \end{array} \right) \, .
\end{equation}
Thus, the scalar product of 4-vectors is:
\begin{eqnarray}
 A\cdot B =A_{\mu}B^{\mu}=g_{\mu\nu}A^{\nu}B^{\mu}=A^0B^0-A^1B^1-A^2B^2-A^3B^3, 
\end{eqnarray}
with
\begin{eqnarray}
 A^{\mu}\equiv(A^0,A^1,A^2,A^3)\equiv(A^0,{\bf a})
\end{eqnarray}
$A^0$ is the temporal component and $A^i$, $i=1,2,3$, the spatial components.
$A_\mu$ is obtained by changing the sing of the spatial components of $A^\mu$.\\

\textbf{Dirac Plane Waves (in coordinate space):}
\begin{itemize}
 \item Particles:
\begin{eqnarray}
 \Psi^{(+)}(X)=\sqrt{\frac{M}{E\ V}} u({\bf p},s)\ e^{-i P\cdot X}
\end{eqnarray}
 
 \item Antiparticles:
\begin{eqnarray}
 \Psi^{(-)}(X)=\sqrt{\frac{M}{E\ V}} v({\bf p},s)\ e^{i P\cdot X}
\end{eqnarray}
\end{itemize}
$V$ is the volume containing the wave function, $M$ and $E$ are energy and mass. ${\bf p}$ is the 3-momentum and $s$ the spin projection.\\

{\bf Dirac Spinors:}
\begin{itemize}
 \item Particles:
\begin{eqnarray}
u({\bf p},s)=\sqrt{\frac{E+M}{2M}} \left(\begin{array}{c}
 \chi_s\\
\dfrac{{\bs \sigma}\cdot{\bf p}}{E+M}\chi_s\end{array}\right) \, ,
\end{eqnarray}
 \item Antiparticles:
\begin{eqnarray}
v({\bf p},s)=\sqrt{\frac{E+M}{2M}} \left(\begin{array}{c}
 \dfrac{{\bs \sigma}\cdot{\bf p}}{E+M}\chi_{-s}\\ 
\chi_{-s}
\end{array}\right) \, ,
\end{eqnarray}
\end{itemize}
where $E=\sqrt{M^2+p^2}$, $\chi_s$ is the bispinor defining the spin projection and ${\bs \sigma}$ are the Pauli matrices. \\

The {\bf normalization} is:
\begin{eqnarray}
 \bar{u}({\bf p},s)u({\bf p},s)=1\,,\,\,\,\,\,\,\,\,\,
 \bar{v}({\bf p},s)v({\bf p},s)=-1\,,
\end{eqnarray}
where $\bar{u}({\bf p},s)=u^\dagger({\bf p},s)\gamma^0$.

One gets:
\ba
    u^\dagger({\bf p},s)u({\bf p},s)= E/M\,.
\ea\\

These spinors are solutions of the {\bf free Dirac equations}: 
\begin{eqnarray}
 (\pslash -M) u({\bf p},s)=0\,,\,\,\,\,\,\,\,\,\,
 (\pslash +M) v({\bf p},s)=0\,.
\end{eqnarray}

{\bf Other useful relations:}

For a given spin projection $s$:
\begin{itemize}
 \item Particles:
\begin{eqnarray}
u_{\alpha}({\bf p},s)\bar{u}_{\beta}({\bf p},s) = 
\left[\frac{\displaystyle{\not} P + M}{2M}
\frac{\munit+\gamma_5\displaystyle{\not} S}{2}\right]_{\alpha\beta} 
\end{eqnarray}

\item Antiparticles:
\begin{eqnarray}
v_{\alpha}({\bf p},s)\bar{v}_{\beta}({\bf p},s) = 
\left[\frac{\displaystyle{\not} P - M}{2M}
\frac{\munit+\gamma_5\displaystyle{\not} S}{2}\right]_{\alpha\beta} \, .
\end{eqnarray}
\end{itemize}

Summing on the spin (up and down):
\begin{itemize}
 \item Particles:
\begin{eqnarray}
\sum_s u_{\alpha}({\bf p},s)\bar{u}_{\beta}({\bf p},s) = 
\left[\frac{\displaystyle{\not} P + M}{2M}\right]_{\alpha\beta} 
\end{eqnarray}

\item Antiparticles:
\begin{eqnarray}
\sum_s v_{\alpha}({\bf p},s)\bar{v}_{\beta}({\bf p},s) = 
\left[\frac{\displaystyle{\not} P - M}{2M}\right]_{\alpha\beta} \, .
\end{eqnarray}
\end{itemize}

Closure relation:
\begin{eqnarray}
 \sum_s[u_{\alpha}({\bf p},s)\bar{u}_{\beta}({\bf p},s)-
 v_{\alpha}({\bf p},s)\bar{v}_{\beta}({\bf p},s)]=\delta_{\alpha\beta}
 \label{completitud}
\end{eqnarray}\\

\textbf{Fourier transform of the fermion wave functions:}

\begin{eqnarray}
 \Psi({\bf p})=\frac{1}{(2\pi)^{3/2}}\int d{\bf x}\ 
 \Psi({\bf x})e^{-i{\bf x}\cdot{\bf p}}\, ,
\end{eqnarray}
and
\begin{eqnarray}
 \Psi({\bf x})=\frac{1}{(2\pi)^{3/2}}\int d{\bf p}\ 
 \Psi({\bf p})e^{i{\bf x}\cdot{\bf p}}\, ,
\end{eqnarray}
From those results, one gets the following normalization:
\begin{eqnarray}
 \int d{\bf x}\ |\Psi({\bf x})|^2=\int d{\bf p}\ |\Psi({\bf p})|^2=1.
\end{eqnarray}\\

{\bf Dirac delta:} 
\ba
    \int dX\, e^{i(P'-P)\cdot X} = (2\pi)^4\delta^4(P'-P)\,.\label{Delta}
\ea
If $P'=P$, then $(2\pi)^4\delta^4(P'-P) \longrightarrow V \cdot T$.\\

The squared of Dirac delta is (see pg 85 and 108 in~\cite{GreinerQED}):
\ba
    [(2\pi)^4\delta^4(P'-P)]^2 = T V\, (2\pi)^4\delta^4(P'-P)\,.
\ea

Another useful property is:
\ba
    \left. \delta[f(x)]=\sum\dfrac{\delta(x-x_i)}{|\partial f(x)/\partial x|}\right|_{x=x_i}\,,\label{delta-prop}
\ea
where $x_i$ is such that $f(x_i)=0$.\\

{\bf Fully antisymmetric Levi-Civit\'a tensor}, 
$\epsilon^{\alpha\beta\gamma\delta}$: $\epsilon^{0123} = +1$ and $\epsilon_{0123} = -1$.\\

\textbf{Other definitions:}
\begin{eqnarray}
\gamma^5=i\gamma^0\gamma^1\gamma^2\gamma^3 \, .
\end{eqnarray}

Antisymmetric tensor $\sigma^{\mu\nu}$:
\begin{eqnarray}
 \sigma^{\mu\nu}=\frac{i}{2}\left[\gamma^{\mu},\gamma^{\nu}\right] 
 = i\left(\gamma^\mu\gamma^\nu -g^{\mu\nu}\right) \, .
\end{eqnarray}

Anticonmutation relations of the Dirac matrices:
\begin{eqnarray}
 \lbrace \gamma^{\mu},\gamma^{\nu}\rbrace = 2g^{\mu\nu}\,,\,\,\,\,\,\,\,\,\,
\lbrace \gamma^5,\gamma^{\mu}\rbrace = 0 \, .
\end{eqnarray}

\section{Feynman rules in coordinate space}\label{Feyn-rules}

To compute the amplitude of a given Feynman diagram, associated to a given scattering process, one can use the following Feynman rules in coordinate space:
\begin{enumerate}
  \item Each vertex: $i\int{\de^4X}{\cal \bar{L}}$, where ${\cal \bar{L}}$ is the Lagrangian without the fields.
  \item Particle propagating from X to Y with 4-momentum $P^\mu$:
  \begin{itemize}
    \item W boson: $$D_W(X-Y) = \int{\frac{\de^4P}{(2\pi)^4}}\frac{-ig_{\mu\nu}}{P^2-M_W^2}e^{iP\cdot(X-Y)}\,.$$
    \item Photon: $$D_\gamma(X-Y) = \int{\frac{\de^4P}{(2\pi)^4}}\frac{-i g_{\mu\nu}}{P^2}e^{iP\cdot(X-Y)}\,.$$
  \end{itemize}
  \item External particle with 4-momentum $P^\mu=(E,\np)$: Incoming fermion: $\Psi(X)$, outgoing fermion $\Psib(X)$.
  %

\end{enumerate}

\section{Lagrangian densities: Couplings to external fields}

The Standard Model of Particle Physics gives us the full Lagrangian density ${\cal L}$ which describes all particles and their interactions. 
Here, we are interested only in the interaction of fermions with external fields (bosons).

\subsection{Photon-lepton-lepton vertex}
The coupling of an electron to a photon is given by: 
\ba
    {\cal L}_{\gamma ee} = e\psib_e\gamma^\mu\psi_e\ A_\mu\,,
\ea
where $\psib_e$ is the electron field, $A_\mu$ the photon field and $e$ is the coupling constant (electron charge). For chargeless leptons (neutrinos), the coupling is zero.

\subsection{Photon-nucleon-nucleon vertex} 

If nucleons are considered as point-like particles, the expression is the same as for the electron but changing the sign of the coupling 
\ba
    {\cal L}_{\gamma NN} &=& -e\psib_p\gamma^\mu\psi_p\ A_\mu\,,
\ea
notice that there is no coupling to the neutron because the neutron net charge is zero.

The internal structure of the nucleons is accounted for by replacing the simple $\gamma^\mu$ by a generic operator $\hat{\Gamma}_N^\mu$, which has to fulfill some symmetries and is discussed elsewhere:
\ba
    {\cal L}_{\gamma NN} = -e(\psib_p \hat{\Gamma}_p^\mu\psi_p 
			  + \psib_n \hat{\Gamma}_n^\mu\psi_n)\ A_\mu\,,\label{LphNN}
\ea
with $\hat{\Gamma}_{p,n}^\mu$ the proton (p) and neutron (n) vector current operator described in Appendix~\ref{app:structure}.
They are both functions of the squared 4-momentum transfer $Q^2$, with $Q=P_f-P_i$ being the difference between the initial and final 4-momentum of the nucleon. Notice that now, the photon can couple to the neutron, and by convention one takes the same sign of the coupling as for the proton.

\subsection{$W$ boson-neutrino-lepton vertex}
The coupling of the W boson to a neutrino reads:
\ba
{\cal L}_{W {\nu_\ell} \ell} = \frac{-g}{2\sqrt2}\Big[ 
        \psib_{\ell}\gamma^\mu(1-\gamma^5)\psi_{\nu_\ell} W_\mu^- 
	  + \psib_{\nu_\ell}\gamma^\mu(1-\gamma^5)\psi_\ell   W_\mu^+\Big]\,.
\ea
For antiparticles, one has:
\ba
{\cal L}_{W {\bar\nu_\ell} \bar\ell} &=& \frac{-g}{2\sqrt2}\Big[ 
	  \psib_{\bar\ell}\gamma^\mu(1+\gamma^5)\psi_{\bar\nu_\ell} W_\mu^+ 
	  + \psib_{\bar\nu_\ell}\gamma^\mu(1+\gamma^5)\psi_{\bar\ell}   W_\mu^-\Big]\,.
\ea
The field $W^+_\mu$ creates a $W^-$ boson or annihilates a $W^+$ boson, therefore the $W^-_\mu$ field creates a $W^+$ boson or annihilates a $W^-$ boson.

\subsection{$W$ boson-nucleon-nucleon vertex}
The coupling of a W boson to a nucleon reads:
\ba
{\cal L}_{W NN} &=& \frac{-g}{2\sqrt2}\Cab\Big[ 
        \psib_p\gamma^\mu(1-\gamma^5)\psi_n W_\mu^+ 
      + \psib_n\gamma^\mu(1-\gamma^5)\psi_p W_\mu^- \Big]\non\\
&\longrightarrow& \frac{-g}{2\sqrt2}\Cab\Big[ 
        \psib_p(\hat{\Gamma}^\mu_V-\hat{\Gamma}^\mu_A)\psi_n W_\mu^+ 
	  + \psib_n(\hat{\Gamma}^\mu_V-\hat{\Gamma}^\mu_A)\psi_p W_\mu^- \Big]\,,\label{LWNN}
\ea
with $\hat{\Gamma}^\mu_V$ and $\hat{\Gamma}^\mu_A$ the vector and axial-vector current operators for the isovector transition given in Appendix~\ref{app:structure}.
$\Cab=0.974$ is the Cabibbo quark-mixing angle.

\section{Internal structure of the nucleon}\label{app:structure}

To understand why the nucleon current operator is what it is, see e.g. discussion starting at page 121 in~\cite{GreinerQED}. There, it is shown that the operator in eq.~\ref{CC2em} is the most general expression for a transition current that fulfills the conditions of Lorentz covariance, Hermiticity, gauge invariance (or current conservation) and invariance under parity transformation. The same arguments stand for the weak operator of eq.~\ref{GammaVA} but, in that case, one does not consider current nor parity conservation, what gives rise to two additional terms: the axial-vector ($\sim\gamma^\mu\gamma^5$) and pseudoscalar ($\sim\gamma^5$) contributions. 
A summary with different prescriptions of the nucleon form factors can be found in Section 3 in Ref.~\cite{Gonzalez-Jimenez13a}.

Two prescriptions are often found in the literature for the vector current operator, namely, CC1 and CC2:
\begin{eqnarray}
\left. \Gamma^{\mu}_{EM}\right|^{p,n}_{CC1}  &=& 
(F_1^{p,n}+F_2^{p,n})\gamma^{\mu}-\frac{F_2^{p,n}}{2M}(P+P_N)^\mu \, ,\label{CC1em}\\
\left. \Gamma^{\mu}_{EM}\right|^{p,n}_{CC2} &=& 
F_1^{p,n}\gamma^{\mu}+i\frac{F_2^{p,n}}{2M}\sigma^{\mu\nu}Q_{\nu}\, .\label{CC2em}
\end{eqnarray}
The superscript $^{p,n}$ means proton (p) or neutron (n). $F_{1,2}=F_{1,2}(Q^2)$ are the elastic nucleon form factors, which are functions of the squared 4-momentum transfer. For these structure functions, both phenomenologic and microscopic models are available in the literature (see Section 3 in Ref.~\cite{Gonzalez-Jimenez13a} for a review).

These two forms of the operator are equivalent for on-shell spinors~\footnote{On-shell means ``on the mass shell'', which actually means that the energy and momentum are correlated by the familiar relation $E^2=p^2+M^2$, $M$ being the invariant mass. It is important to stress that the equivalence of CC1 and CC2 stands only for on-shell particles. For instance, contrary to what happens in a Fermi gas approach, in mean-field based models, the bound nucleons are off-shell: they are described by energy-eigenstate wave functions, i.e., they have well defined energy but the momentum is given by a probability distribution and their masses do not coincide with the free-nucleon mass.}: 
\ba
   \left.  \ubar\Gamma^{\mu}_{EM}\right|_{CC1}\uu =    \left.  \ubar\Gamma^{\mu}_{EM}\right|_{CC2}\uu\label{Gordon}
\ea 
This equation is known as Gordon relation.\\

In the case of charged-current interaction, the current operators read:
\ba
\hat{\Gamma}^\mu_V &=& F_1^V\gamma^\mu + i\frac{F_2^V}{2M}\sigma^{\mu\alpha}Q_\alpha\,,\\
\hat{\Gamma}^\mu_A &=& G_A\gamma^\mu\gamma^5 + F_PQ^\mu\gamma^5\,.\label{GammaVA}
\ea
One has $F_{1,2}^V = F_{1,2}^p-F_{1,2}^n$, with $F_{1,2}^{p,n}$ the proton and neutron electromagnetic form factors (eqs.~\ref{CC1em} and \ref{CC2em}). $G_A$ is the axial-vector isovector form factor which is usually described by a dipole form $G_A=g_A/(1-Q^2/M_A^2)^2$. The pseudoscalar form factor is related to the axial-vector one by PCAC $F_P=G_A\,2M/(|Q^2|+m_\pi^2)$. For more details see e.g. Section 3 in Ref.~\cite{Gonzalez-Jimenez13a}.

\section{Kinematics for elastic lepton-nucleon scattering with massive leptons}\label{full-kinematics}

If we do not consider the ultrarelativistic limit, instead of eq.~\ref{Kinelastic}, one finds:
\ba
  f(\varepsilon_f)=-\varepsilon_i + \varepsilon_f +\sqrt{k_i^2 + k_f^2 - 2k_i k_f\cos\theta_f + M^2} - M=0\,.\label{Kinelastic-full}
\ea
The solution for $\varepsilon_f$ is:
\ba
  \varepsilon_f = \frac{AB \pm \sqrt{(AB)^2 - (B^2-C^2)(A^2+C^2m_f^2)} }{ B^2-C^2}\,
\ea
with $+$ for $\cos\theta_f>0$ and $-$ for $\cos\theta_f\leq 0$, and:
\ba
  A &=& \varepsilon_iM + \frac{m_i^2+m_f^2}{2}\,,\non\\
  B &=& \varepsilon_i + M\,,\non\\
  C &=& k_i\cos\theta_f\,.\non
\ea

The recoil factor $f_{rec}$ that appears in the cross section is given by 
\ba
  f_{rec} = \left|\frac{\partial f(\varepsilon_f)}{\partial\varepsilon_f}\right| 
  = \left| 1 + \frac{\partial E_N}{\partial k_f}\frac{\partial k_f}{\partial \varepsilon_f} \right| 
  = \left| 1 + \frac{k_f-k_i\cos\theta_f}{E_N} \frac{\varepsilon_f}{k_f} \right|.
\ea

\section{Jacobian for the missing energy distributions with nuclear recoil}\label{app:Jacobian}

From energy-momentum conservation we have:
\ba
    \sqrt{p_N^2+M_N^2} &=& M_A + \omega - \sqrt{ q^2 + p_N^2 -2qp_N\cos\theta_{qN} + (E_m - M_N + M_A)^2 }\,,
\ea
where we have used the kinematic relations:
\ba
    E_B^2 &=& p_B^2 + (M_B)^2\,,\non\\
    M_B &=& E_m - M_N + M_A\,,\non\\
    p_B^2 &=& q^2 + p_N^2 -2qp_N\cos\theta_{qN}\,.\non
\ea
Thus, the absolute value of the Jacobian reads
\ba
   {\cal J}=\left|\frac{dE_m}{dp_N}\right| = \frac{E_B}{M_B}\left|\frac{p_N}{E_N} + \frac{p_N-q\cos\theta_{qN}}{E_B}\right|\,.\label{Jacobian}
\ea
It is easy to show that the Jacobian~\ref{Jacobian} is equivalent to eq.~\ref{Jacobian_easy} if one neglects the recoil.

\section{Normalization of the momentum distribution in the  lRFG model}\label{app:norma}

The momentum distribution in the local Fermi gas model reads:
\ba
 n_{lRFG}(p) = \int\de\nr\, \frac{\rho(r)}{{\cal N}} \left[{\cal N}\frac{\Theta\left[p_F(r)-p\right]}{\frac{4}{3}\pi p_F(r)^3}\right].\label{norma0}
\ea
We are going to demonstrate that 
$\int\de\np\, n_{lRFG}(p) = {\cal N}$.
Using eq.~\ref{norma0}, we get:
\ba
\int\de\np\, n_{lRFG}(p) &=& \int\de\np\,\int\de\nr\, 
 \frac{\rho(r)}{{\cal N}} \left[{\cal N}\frac{\Theta\left[p_F(r)-p\right]}{\frac{4}{3}\pi p_F(r)^3}\right]\non\\
 &=& (4\pi)^2 \int\de r r^2 \frac{\rho(r)}{\frac{4}{3}\pi p_F(r)^3} 
 \int\de p p^2\, \Theta\left[p_F(r)-p\right]\non\\
 &=& (4\pi)^2 \int\de r r^2 \frac{\rho(r)}{\frac{4}{3}\pi p_F(r)^3} 
 \int_0^{p_F(r)}\de p p^2\non\\
 &=& (4\pi)^2 \int\de r r^2 \frac{\rho(r)}{\frac{4}{3}\pi p_F(r)^3} 
 \frac{p_F(r)^3}{3}\non\\
 &=& 4\pi \int\de r r^2 \rho(r) = {\cal N}\,.\,\,\,\,\,\,\text{QED}
\ea

\section{Analytical expressions of the nuclear density}\label{app:NuclearDensity}

For convenience, we have used analytical expressions of the nuclear density. We take them directly from~\cite{LeitnerThesis}, they read:
\ba
  \rho_{n,p}(r) = \rho_0\left[1+a_{n,p}\left(r/R_{n,p}\right)^2\right] \exp\left[-(r/R_{n,p})^2\right]\,.\label{rho_analy}
\ea
The parameters are:
\begin{align}
\text{For $^{12}$C: }&\,  
R_p=1.692\text{ fm}\,,\,\,\,\, R_n=R_p\,,\,\,\,\, a_p=1.082\,,\,\,\,\, a_n=a_p\,.\label{c12_param}\\
\text{For $^{16}$O: }&\,  
R_p=1.833\text{ fm}\,,\,\,\,\, R_n=1.815\text{ fm}\,,\,\,\,\, a_p=1.544\,,\,\,\,\, a_n=1.529\,.\label{o16_param}
\end{align}
The parameter $\rho_0$ is defined such that it gives the right normalization. It is possible to analytically solve the integral over $r$, obtaining:
\ba
{\cal N} 
= \int\de\nr\rho_{n,p}(r) 
= 4\pi \rho_0 \frac{\sqrt\pi}{8}(3a_{n,p}+2)R_{n,p}^3\,.
\ea
Hence, we can define $\rho_0$ in terms of the parameters
\ba
 \rho_0 = \frac{2{\cal N}}{\pi\sqrt\pi (3a_{n,p}+2)R_{n,p}^3} 
\ea

Another advantage of the analytical formula is that we can find the maximum Fermi momentum. Looking at eq.~\ref{pFr}, it is clear that the maximum of $p_F(r)$ will occur at the maximum of $\rho(r)$. The maximum is given by $\frac{\de}{\de r}\rho(r)=0$, i.e.:
\ba
 \frac{\de}{\de r} \left[1+a_{n,p}\left(r/R_{n,p}\right)^2\right] \exp\left[-(r/R_{n,p})^2\right] = 2r\exp\left(-(r/R_p)^2\right)(a_p(r^2-R_p^2)+R_p^2) = 0\,. 
\ea
This equations has two solutions for $r$, the trivial $r=0$, which in general is not the maximum, and 
\ba
 r_{max} = \sqrt{\frac{R_p^2}{a_p}(a_p-1)}\,.
\ea
So, the maximum Fermi momentum is obtained plugging this $r_{max}$ into eq.~\ref{pFr}. 

For $^{12}$C and the set of parameters in eq.~\ref{c12_param} we get:
\ba
  r_{max} = 0.466\text{ fm}\,,\,\,\,
  \rho(r_{max}) = 0.0851\text{ fm}^{-3}\,,\,\,\,
  p_F(r_{max}) = 268.5\text{ MeV}\,.
\ea

\newpage

\bibliographystyle{apsrev4-1}
\bibliography{bibliography}

\end{document}